\documentclass[prb,amsmath,amssymb,superscriptaddress,groupaddress,showpacs,
twocolumn,reprint]{revtex4}

\usepackage{graphicx}
\usepackage[utf8]{inputenc}
\usepackage{color}
\usepackage{siunitx}
\usepackage{braket}
\usepackage{amsfonts}
\usepackage{amssymb}
\usepackage{amsmath}
\usepackage{bbold}
\usepackage{booktabs}

\newcommand{\dan}[1]{\operatorname{\hat d}_{#1}}
\newcommand{\dddag}[1]{\operatorname{\hat d}^{\dagger}_{#1}}
\newcommand{\cdag}[1]{\operatorname{\hat c}^{\dagger}_{#1}}
\newcommand{\can}[1]{\operatorname{\hat c}_{#1}}

\newcommand{\tr}[1]{\operatorname{Tr}_{#1}  }
\newcommand{\ketbra}[1]{ \ket{#1}\bra{#1}  }

\newcommand{\ihbar}{\ensuremath{\frac{i}{\hbar}}}

\newcommand{\Deltast}{\Delta_{\textrm{\small{ex}}} }

\newcommand{\up}{\uparrow}
\newcommand{\down}{\downarrow}
\newcommand{\Eq}[1]{Eq.~(#1)}

\newcommand{\mc}[1]{\mathcal{#1}}

\begin{document}

\title{Double island Coulomb blockade in (Ga,Mn)As-nanoconstrictions}

\author{S.~Geißler}
\affiliation{Institute for Experimental
Physics, University of Regensburg, 93040 Regensburg, Germany}

\author{S.~Pfaller}
\email{sebastian1.pfaller@ur.de}
\affiliation{Institute for Theoretical
Physics, University of Regensburg, 93040 Regensburg, Germany}

\author{M. Utz}
\affiliation{Institute for Experimental
Physics, University of Regensburg, 93040 Regensburg, Germany}

\author{D. Bougeard}
\affiliation{Institute for Experimental
Physics, University of Regensburg, 93040 Regensburg, Germany}

\author{A.~Donarini}
\affiliation{Institute for Theoretical
Physics, University of Regensburg, 93040 Regensburg, Germany}

\author{M.~Grifoni}
\affiliation{Institute for Theoretical
Physics, University of Regensburg, 93040 Regensburg, Germany}

\author{D.~Weiss}
\affiliation{Institute for Experimental
Physics, University of Regensburg, 93040 Regensburg, Germany}

\date{\today}

\begin{abstract}
We report on a systematic study of the Coulomb blockade effects in nanofabricated
narrow constrictions in thin (Ga,Mn)As films. Different low-temperature transport regimes have been observed
for decreasing constriction sizes: the ohmic, the
single electron tunnelling (SET) and a completely insulating regime.
In the SET, complex stability diagrams with nested Coulomb diamonds and anomalous
conductance suppression in the vicinity of charge degeneracy points have been observed.
We rationalize these observations in the SET with a double ferromagnetic island model coupled to ferromagnetic leads.
Its transport characteristics are analyzed in terms of a modified orthodox theory of Coulomb blockade
which takes into account the energy dependence of the density of states in the
metallic islands.
\end{abstract}


\maketitle

\section{Introduction}

(Ga,Mn)As, discovered by Ohno et al. \cite{Ohno96}  nearly two decades ago, is by now
the best studied ferromagnetic semiconductor
\cite{DietlOhno14,Jungwirth2006,Sato2010}.
An interesting aspect of this material are large magnetoresistance effects
which were discovered in nanofabricated narrow constrictions in thin (Ga,Mn)As
films \cite{Ruster2003,Giddings2005,Schlapps06,Ciorga2007,Pappert2007,Wunderlich2006}.
While the effects were initially interpreted in terms of the tunneling
magnetoresistance (TMR) \cite{Ruster2003} and tunneling anisotropic
magnetoresistance (TAMR) \cite{Giddings2005}, it was proven later that the interplay
with Coulomb blockade is also relevant in narrow (Ga,Mn)As constrictions \cite{Wunderlich2006,Schlapps2009}.
The origin of this Coulomb blockade anisotropic magnetoresistance (CBAMR) effect are substantial
nanoscale fluctuations in the hole density \cite{DietlOhno14} forming puddles
of high hole density separated by low conducting regions. (Ga,Mn)As is known to
be a strongly disordered material. Its hole density is  close to
the metal-insulator-transition. Little variations in the hole density caused
by local potential fluctuations can lead to an intrinsic structure consisting of
metallic islands separated by insulating areas. It was shown that the
magnetoresistance depends, in the presence of Coulomb blockade, not only on an
applied gate voltage but can also be tuned by changing the direction of the
applied magnetic field \cite{Wunderlich2006,Schlapps2009}. The latter results
from the dependence of the Fermi energy on changes in the magnetization 
$\delta \vec{M}$ and was modeled phenomenologically by Wunderlich et al.
\cite{Wunderlich2006}. If transport occurs through a narrow nanoconstriction,
single electron tunneling (SET) between islands of high carrier density becomes
relevant. Thus it is not surprising that the bias and temperature dependence
of the magnetoresistance for different magnetization directions could be fitted
with a model for granular metals in which metallic islands are separated by
insulating regions \cite{Schlapps2009}. Because of the nanoscale size of the
involved "metallic" islands, the Coulomb-charging energy $U$ is the dominating
energy for transport across the nanoconstriction at low temperatures and small
bias voltages $V_{\rm b}$. Since usually more than one island is involved in transport,
Coulomb blockade diamonds, where the resistance is plotted as a function of
both bias and gate voltage, revealed a very complex and irregular pattern. Up
to now a detailed experimental and theoretical analysis of the Coulomb blockade
effects in (Ga,Mn)As nanoconstrictions in the single-electron-transistor regime is
still missing.

The aim of this work is a systematic study of the Coulomb blockade effects in nanofabricated
narrow constrictions in thin (Ga,Mn)As films. By means of a two step
electron beam lithography (EBL) technique we fabricated well
defined nanoconstrictions (NC) of different sizes. Depending
on channel width and length, for a specific material, different low-temperature
transport regimes could be observed, namely the ohmic regime, the
single electron tunnelling regime and a completely insulating regime.
In the SET regime, complex stability diagrams with nested Coulomb diamonds and anomalous
conductance suppression in the vicinity of charge degeneracy points have been observed.
In order to understand these observations we propose, for a specific nanoconstriction,
a model consisting of two ferromagnetic islands coupled to ferromagnetic leads.
We study its transport characteristics within a modified orthodox theory of Coulomb blockade
which takes into account the energy dependence of the density of states in the
metallic islands.

The paper is structured as follows: Sect.~\ref{sect:sample fab} explains the
fabrication process of the samples. In Sect.~\ref{sect:measurement
setup} the measurement setup is presented. The next section,
Sect.~\ref{sect:results} summarizes the results of the measurements, giving a first
interpretation in terms of a double island structure within a classical orthodox
model of Coulomb blockade \cite{Averin1986,Averin1991,Grabert1991, Grabert1992, Kouwenhoven1997, Barnas2008}. In Sect.~\ref{sect:theoretical
modeling}, we present the details of the ferromagnetic double island model, study its transport
characteristics and make a direct comparison with the experimental results in
Sect.~\ref{sect:comparison}.
Conclusions are drawn in Sect.~\ref{sect:conclusions}.

\section{Sample fabrication}
\label{sect:sample fab}

Our NC-devices were fabricated in a top-down approach starting from a
(Ga,Mn)As-layer with a Mn-content of approximately 5$\%$. The (Ga,Mn)As layer
we used had a thickness of $15\mathrm{nm}$ and was grown by low-temperature
molecular beam epitaxy on top of a (001)-GaAs substrate. In contrast to the
experiments of
Schlapps et al. \cite{Schlapps2009} we used as-grown (Ga,Mn)As samples without
additional annealing before the sample preparation. First of all, we defined
contact-pads for the source- and drain-contacts as well as alignment-marks for
the nanopatterning. This was done using optical lithography followed by thermal
evaporation of 10\,nm Ti and 90\,nm Au  in a standard
lift-off technique. After that,
the NC was defined by means of EBL and
subsequent chemically assisted ion-beam etching using Cl$_2$. A two-step
EBL-process, which allows a precise control of the geometry of the nanocontact and a
reliable processing, was developed and is described in Appendix~\ref{app:sample_fab}

\begin{figure}[h!]
 \includegraphics[width = 0.6\columnwidth]{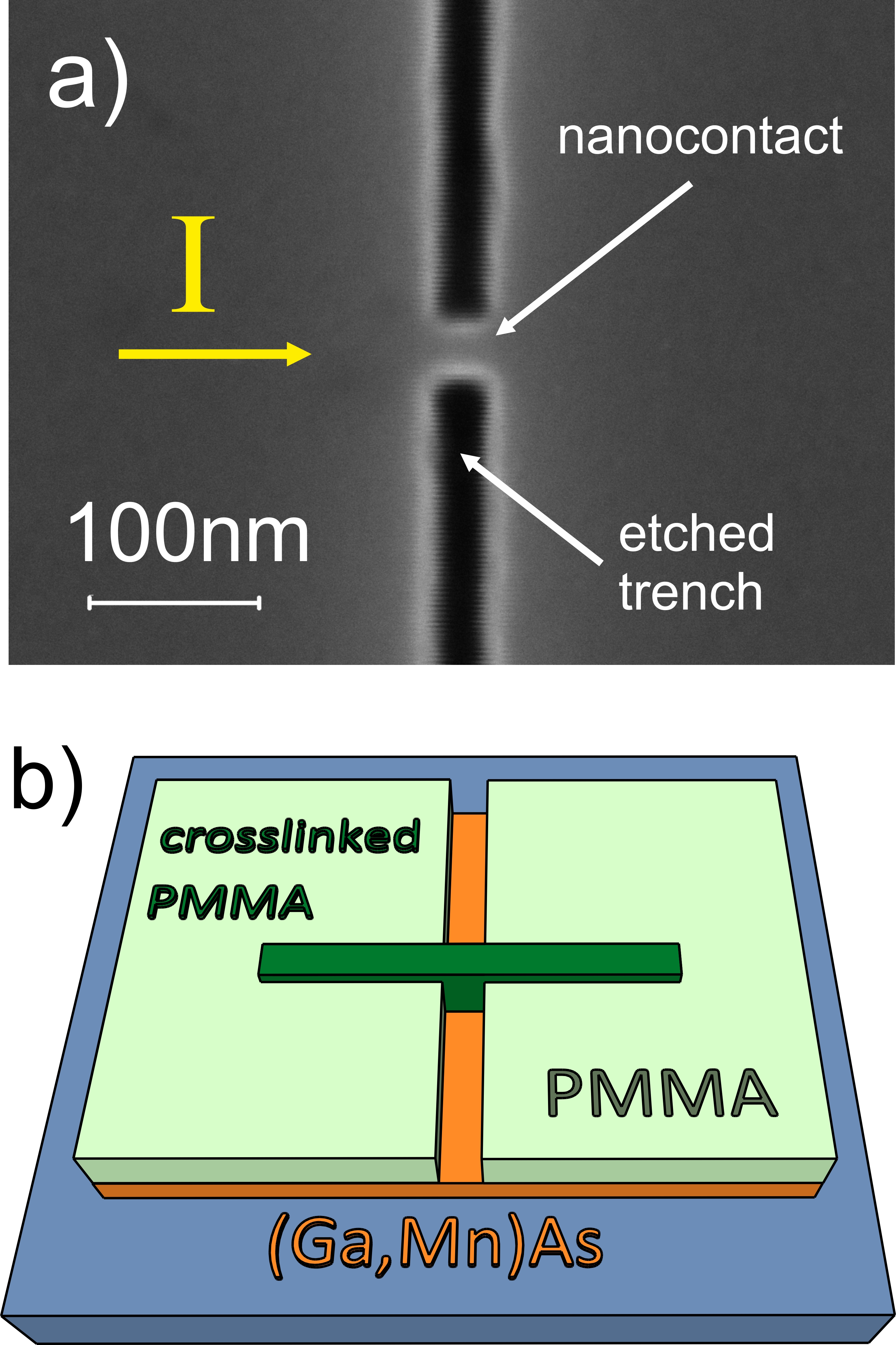}
\caption{a) Schematic of the PMMA-mask (green / light green) defined by a two
step EBL-process for etching the NC-structure into a (Ga,Mn)As-layer (orange) on
top of a semi-insulating GaAs-substrate (blue).
b) Electron micrograph of an NC-device after ion-beam-etching and resist
removal.}
\label{fig:fig1}
\end{figure}

The structure of the PMMA mask, used for the two-step process, is sketched in
Fig.~\ref{fig:fig1}a). It mainly consists of
the crosslinked PMMA-line (dark green) of the first, high dose (30.000 C/cm)
 exposure step as well as of a narrow gap line from the second, usual exposure step,
 which separates the (Ga,Mn)As layer into two parts used as source- and drain-contacts.
 The two parts are connected
with each other only at the NC, where the lines of the two exposure steps cross
each other. This procedure allows us to define the width as well as the length
of the NC by two single lines within independent exposure steps. This completely
rules out the inter-proximity-effect between different exposed elements and
reduces the minimum size of the NC to the smallest achievable linewidth of the
two EBL steps. Compared to a single step process our approach is robust with
respect to minor electron dose variations and thus well reproducible.
Because of this, we were able to fabricate a large number of comparable devices
and even to control the geometry of the NC with a precision of a few
nanometers. Fig.~\ref{fig:fig1}b) shows an electron
micrograph of the central part of a typical NC-device taken after the chemically assisted ion beam
etching and resist removal using a low energy oxygen-plasma. After the nanopatterning we
covered the whole sample with a $30\text{nm}$ thick $\mathrm{Al}_2
\mathrm{O}_3$-layer grown by a low temperature atomic layer deposition process
at a temperature of $90^\circ$\,C. The $\mathrm{Al}_2 \mathrm{O}_3$-layer acts on
 the one hand as the gate-dielectric and on the other hand it protects the tiny
NC against  oxidation. The top-gate contact was defined by optical
lithography and covers not only the NC but also the center part of the whole
device. It consists, similarly to the source- and drain-contacts, of a
$10/90\mathrm{nm}$ thick Ti/Au-stack evaporated thermally and structured using a
standard lift-off technique.

An effective way to influence the transport behavior is to apply
an annealing step after the nanopatterning. We used an
annealing temperature of $150^\circ \mathrm{C}$ and durations from
$30\mathrm{min}$ to $3\mathrm{h}$. The post patterning annealing removes
probably some of the defects induced by chemically assisted ion-beam etching.
This can change an initially insulating sample to one in which Coulomb effects
prevail or even to a conducting one. Annealing before the nanopatterning \cite{Schlapps2009,Edmonds2004}, which
removes defects induced during low temperature molecular beam epitaxy growth,
is less effective than  the post patterning annealing. Hence, the intrinsic structure of the NC is
dominated by defects induced during the nanopatterning rather than by defects
stemming from the low temperature molecular beam epitaxy  growth.

\section{Measurement setup}
\label{sect:measurement setup}

All low temperature measurements presented in this work were carried out at a
temperature of about $25\mathrm{mK}$ using a
$^3\mathrm{He}/^4\mathrm{He}$-dilution fridge,  equipped with a
superconducting coil magnet. In combination with a rotatable sample holder, we
were able to apply magnetic fields up to $19\mathrm{T}$ in any direction
parallel to the sample plane. In order to saturate the magnetization of the
device and to fix its direction, we applied a constant in-plane magnetic field with a
magnitude of $1\mathrm{T}$ along one of the easy axes of the extended (Ga,Mn)As
 layer. This leads to a situation as sketched in
Fig.~\ref{fig:fig5}a). The
electrical transport experiments were carried out in a two terminal setup. We
performed ac and dc measurements simultaneously by applying a dc bias-voltage
$V_{\rm dc}$ modulated with a small oscillating ac component $V_{\rm ac}$. The current $I$
flowing through the device was measured using a current amplifier which
also converts the current into a corresponding voltage signal. The dc
measurement using a digital multimeter provides the well known
$I\text{-}V_{\rm dc}$ characteristic, while the ac measurement using a lock-in
amplifier offers the differential conductance  $G= dI/dV_{\rm ac}$ of the device.
Our device could be tuned additionally by an external dc voltage ($V_{\rm g}$) applied
to the top-gate electrode of the device.

\section{Experimental Results}
\label{sect:results}

\subsection{Room temperature properties}

As mentioned in the introduction, all nanoconstricted (Ga,Mn)As-devices  investigated in previous studies
have shown a rather complex and irregular Coulomb diamond pattern \cite{Wunderlich2006,Schlapps2009}. This has
been explained by assuming that several metallic islands are involved in transport
across the NC. Hence, shrinking the size of the NC should reduce the number of
islands within the NC and bring up a more regular Coulomb diamond pattern.
Looking for such samples, we investigated many different devices with widths and lengths of the NC
ranging from $10\mathrm{nm}$  to $100\mathrm{nm}$. Our experiments revealed
that the transport properties of these devices are very sensitive to the width
$w$ of the NC while its length $L$ has only a minor influence. Wider samples
($w>25\mathrm{nm}$) show a mainly ohmic behavior while the most narrow ones
($w<15\mathrm{nm}$) are fully insulating. Only samples with intermediate
widths of $15-25\text{nm}$ show the typical SET-like behavior, discussed below.
In many cases the room temperature resistance $R_{NC}$ of the nanocontact already 
indicates whether the constriction
is insulating, in the Coulomb blockade regime, or ohmic: For $R_{NC }/R_s$ values
(with the sheet resistance of $R_s \sim 4$\,k$\Omega$ at 4.2\,K) 
between 10 and 15 the constriction was in most cases in the Coulomb blockade regime
for this specific material.
However, similar to the earlier experiments,
all of our SET-like samples, even the shortest and narrowest ones, have shown, on a
first  glance, an irregular Coulomb
diamond pattern. Below we discuss in more detail
transport in the Coulomb blockade regime.

\begin{figure}[h!]
  \includegraphics[width = 0.9\columnwidth]{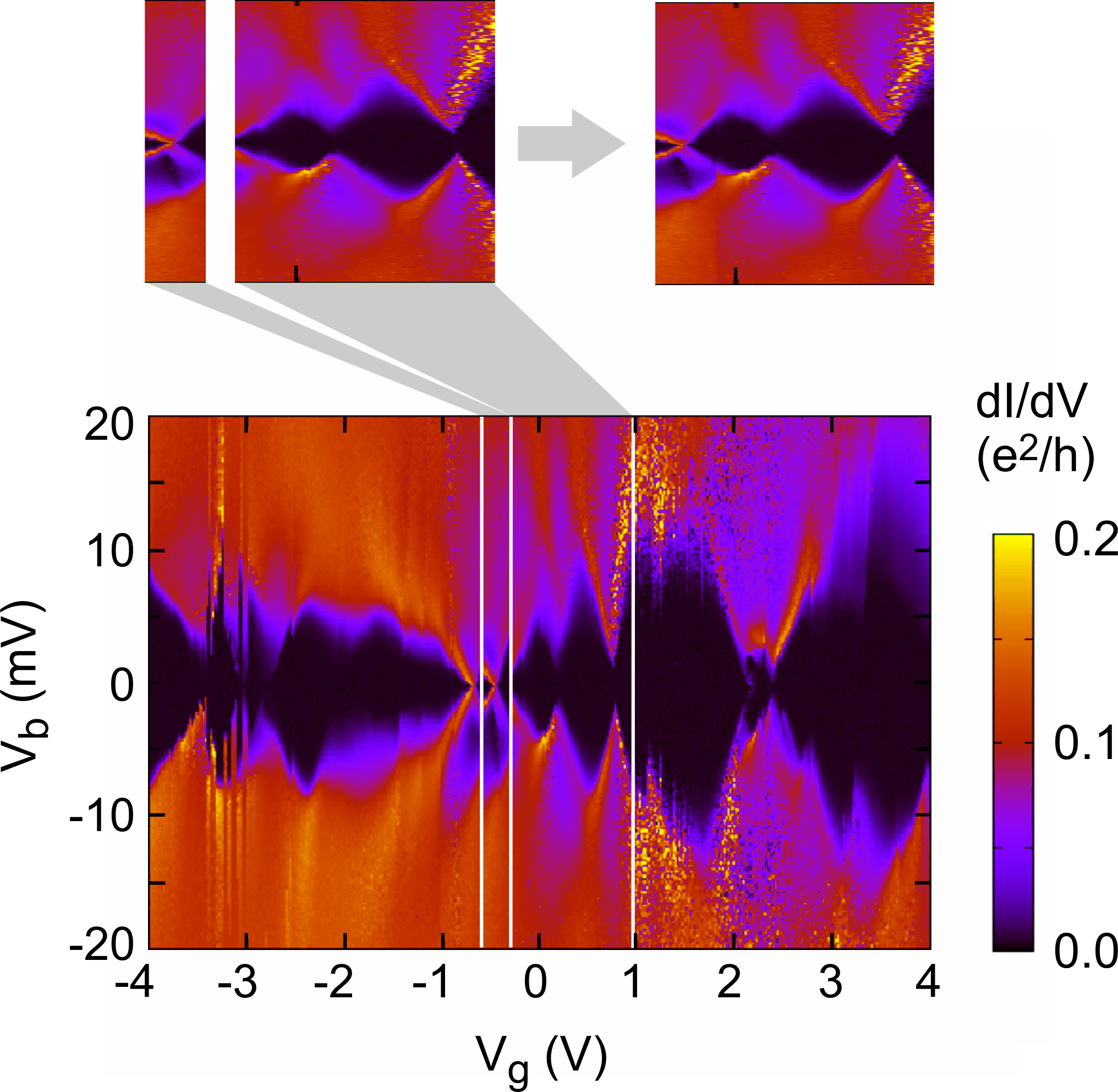}
\caption{ Differential conductance as a function of the
bias- and gate-voltage  of the NC-device in Fig.~\ref{fig:fig1}. The
measurement was done at a temperature of $T= 25$\,mK.
A partial irregular Coulomb diamond pattern with frequently occurring vertical
discontinuities is observed. Three of those discontinuities are marked by white lines. Cutting
the dataset between two of these lines gives an undisturbed segment; stitching
neighboring segments together as described in the text and shown in the upper
inset allows to reconstruct the Coulomb diamond spectrum over a larger gate
voltage range. }
\label{fig:fig2}
\end{figure}

\subsection{Coulomb blockade regime}

\begin{figure}[h!]
  \includegraphics[width = 0.8\columnwidth]{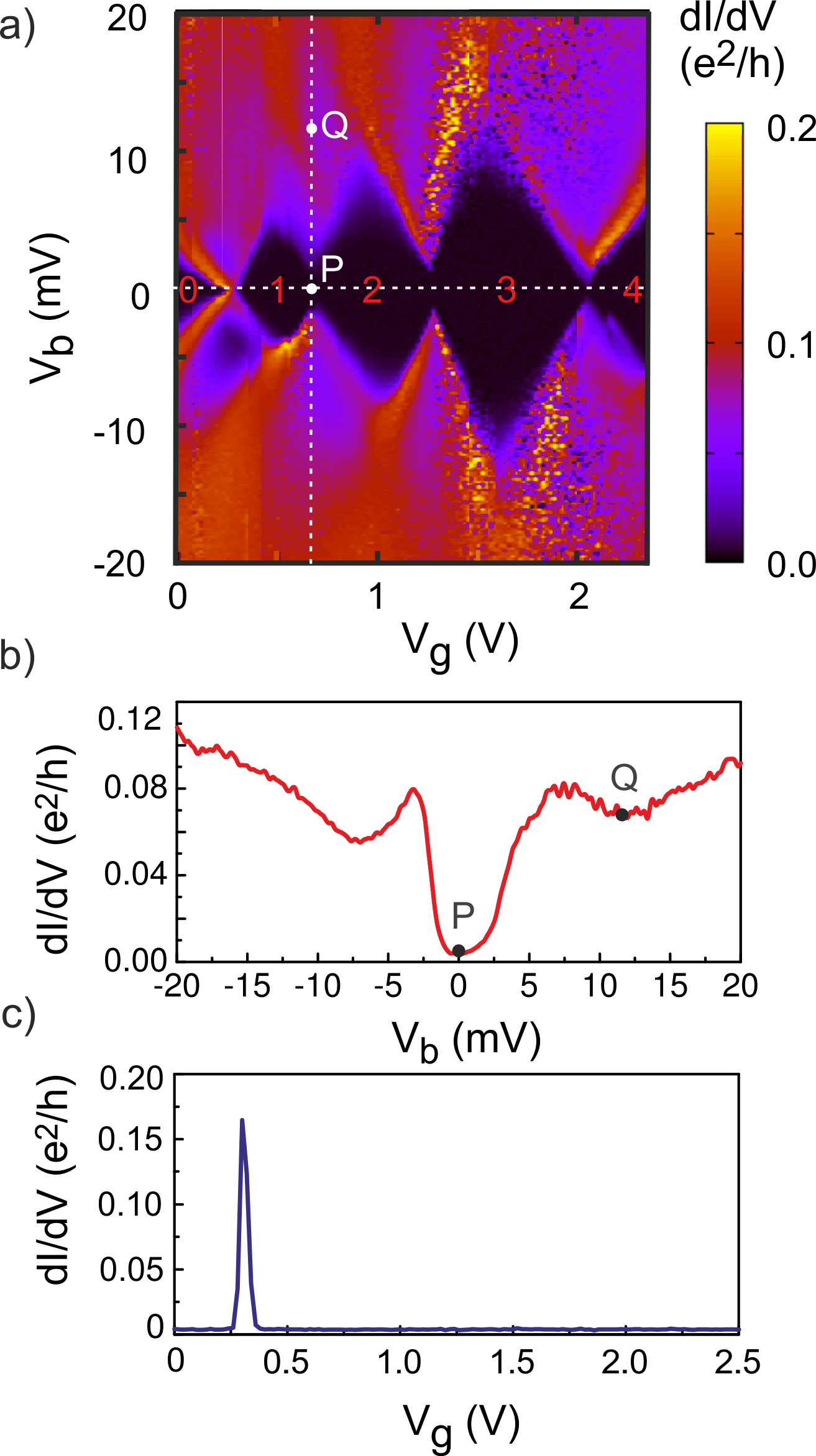}
\caption{
a) Differential conductance of the NC-device of Fig.~\ref{fig:fig2} vs. applied
gate- and bias-voltages after reconstruction. Diamonds labeled 0 to 4 can
clearly be identified. b) Differential conductance as a function of the bias 
voltage corresponding to the vertical dashed line in a). c)
 Conductance at $V_b=0$ as a function of the  gate voltage
corresponding to the horizontal dashed line in a). 
It shows a conductance peak at the 0-1,  and
a blockade at the other charge degeneracy points, including point P. }
\label{fig:fig3}
\end{figure}

In Fig.~\ref{fig:fig2} we present a highly resolved stability diagram of one of 
our
 NC-devices in the SET regime. The first impression is that the Coulomb diamond 
pattern is very
irregular and exhibits frequent vertical discontinuities. Three of them are 
highlighted
by white lines. These abrupt shifts can be assigned to
charging or discharging of local traps in close vicinity to the NC, which, with 
their electrostatic potential,
act as local gates. Their effect can thus be described as an abrupt jump along 
the gate voltage axis.
This observation suggests a method to reconstruct the stability diagrams with
unperturbed Coulomb diamonds. We cut the dataset in
Fig.~\ref{fig:fig2} along the white lines and shift the segments on the
$V_{\rm g}$-axis until the diamonds fit onto each other. An example of this 
procedure
is shown in the top inset of Fig.~\ref{fig:fig2}. In this way we obtain, for 
some
parts of the $V_{\rm g}$-scale, Coulomb diamonds which are essentially cleared 
of potential jumps due to charge fluctuations in local traps. The
dataset displayed in
Fig.~\ref{fig:fig3} has been reconstructed from the data shown in
Fig.~\ref{fig:fig2} and represents the starting point of our more detailed 
analysis.

\begin{figure}[h!]
  \includegraphics[width = 0.8\columnwidth]{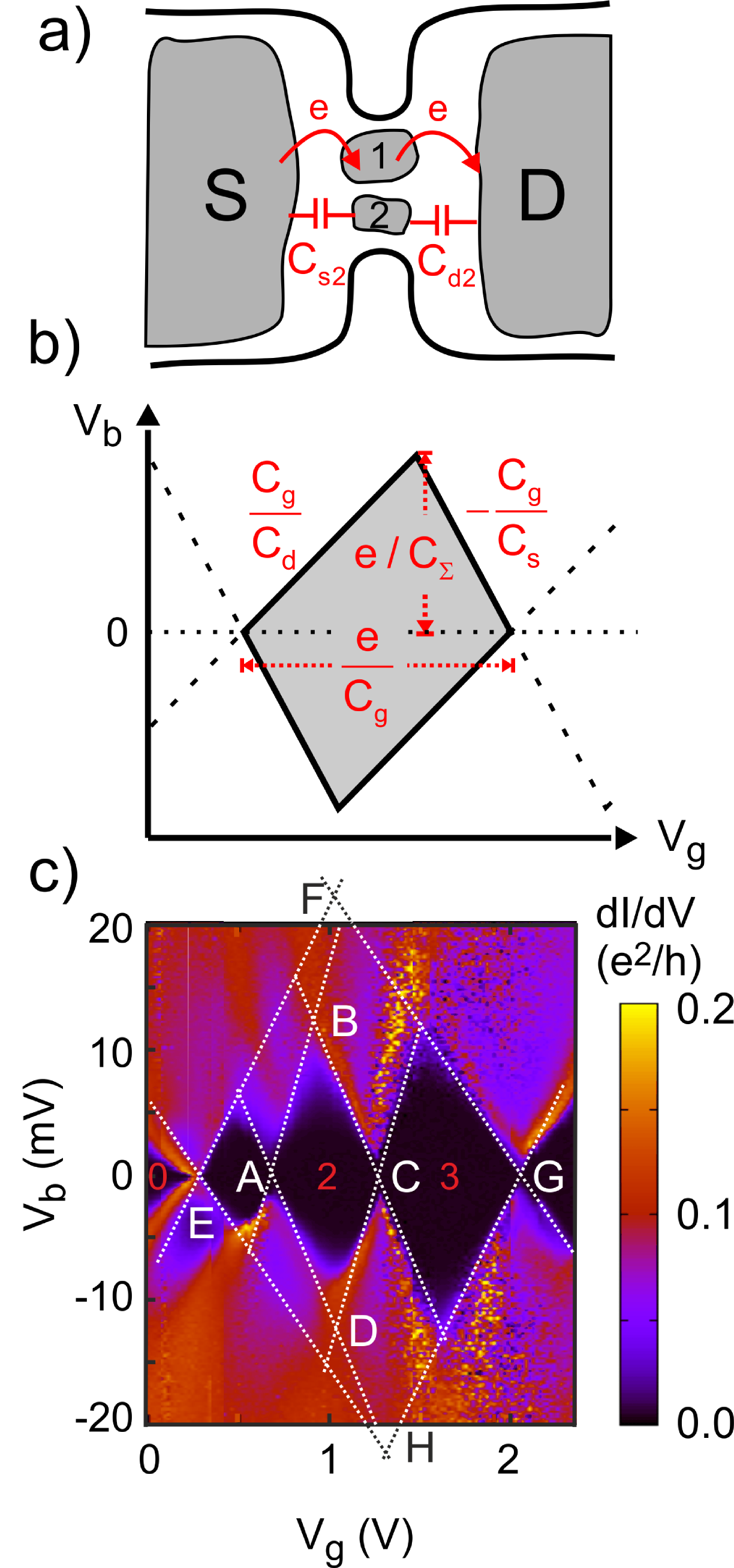}
\caption{a) Schematic of a double island
structure in a parallel configuration. Transport from source to
drain is carried by two subsequent direct tunneling-processes involving only one
of the islands. The two islands are characterized by a
different capacitive coupling to the leads ($C_{{\rm s}i}$,$C_{{\rm d}i}$ ) as 
well as by a
different gate capacitance ($C_{{\rm g}i}$) with $i = 1,2$.
b) Schematic to illustrate the parameter extraction from a regular
Coulomb-diamond in the framework of the orthodox model.
c) The two Coulomb-diamonds (ABCD and EFGH) used to extract the
parameters, are marked by white dotted lines.}
\label{fig:fig4}
\end{figure}

The stability diagram shown in Fig.~\ref{fig:fig3} presents characteristic 
features typical for metallic
single electron transistors\cite{Averin1986, Averin1991,Grabert1991, 
Grabert1992, Kouwenhoven1997} but also several anomalies.
As expected, a series of diamonds of exponentially low differential
conductance (black regions with fixed particle number) are surrounded by ridges 
of high conductance. Moreover, by further increasing
the bias, the differential conductance does not drop to zero, see {\it e.g.} 
Fig.~\ref{fig:fig3}b), allowing to exclude the single particle energy 
quantization
typical for quantum dots. Unexpectedly, though, i) the size and the shape of 
the Coulomb diamonds is not regular, ii) some of the
diamonds are not closing at zero bias ({\it e.g.} corners between diamond 1 
and 2 or between diamond 2 and 3 as seen from the gate trace in
Fig.~\ref{fig:fig3}c)).

Concerning the first anomaly, it is striking that all the diamonds exhibit an
individual height as well as an individual width. Additionally the diamond 
labeled 1 and
the diamond labeled 3 are asymmetric: according to the classical orthodox 
theory \cite{Averin1986}, one would
expect that all Coulomb diamonds associated to a single island have the same size and
shape, and that opposing
edges of a Coulomb diamond were parallel.
In the orthodox picture the two different slopes of a Coulomb diamond are
related to the capacitive coupling of the island to the source- ($C_{\rm s}$) 
and
drain-leads ($C_{\rm d}$), as well as to the gate electrode ($C_{\rm g}$). 
Assuming $C_{\rm g} \ll
C_{\rm s,d}$, the slope of the source-line is given by $C_{\rm g}/C_{\rm d}$  
while the slope of the
drain-line is given by $-C_{\rm g}/C_{\rm s}$ (see Fig.~\ref{fig:fig4}). In our 
case only
the diamond numbered 2 has parallel source and drain lines. The diamonds
labeled 1 and 3, however, exhibit four different slopes, so that we would 
extract from each two
different values for $C_{\rm s}$ and $C_{\rm d}$  or two different values for 
$C_{\rm g}$,
respectively. This suggests that our NC consists actually  of two metallic 
islands
producing a set of nested diamonds.

Fig.~\ref{fig:fig4}a) shows a simple schematic to illustrate our interpretation:
the two islands are arranged in parallel, so that an electron can tunnel from
the source-lead directly to each of the two islands and from there in
a subsequent tunneling  process directly to the drain-lead.
By taking into account the slopes of the diamond edges as well as the distance
between neighboring charge degeneracy points we can obtain two different sets of
parameters ($C_{\rm s}$, $C_{\rm d}$, $C_{\rm g}$) from our experimental
data. Each set of parameters characterizes one of the two islands. One set can
be extracted from the regularly shaped diamond 2. For the other one, we have to
reconstruct a second regular Coulomb diamond by extending the outer edges of
diamond 1 and 3 until they cross each other, see Fig.~\ref{fig:fig4}c). The 
extracted
parameters  are summarized in Table~\ref{tab:tab1}.
Our analysis is limited to certain gate voltage ranges. We attribute this 
limitation to
possible differences in the shape and even in the number of participating
island associated to different gate voltage regions.
Nevertheless, the simple orthodox model gives already a satisfactory agreement 
between experimental
and theoretical $dI/dV_{\rm b}$-stability diagrams and suggests that transport, 
in this gate voltage range,
occurs primarily in parallel
across two islands of different size in the reconstructed gate voltage segments.
However, the model  presented  so far can not account for the second
anomaly, 
{\it i.e.}
a pronounced transport blocking observed in the vicinity of
the charge degeneracy point between the diamonds 1-2 and 2-3, see also 
Fig.~\ref{fig:fig2}b).
On the other hand, the gap is not present at the charge
degeneracy point 0-1 and is barely visible at  3-4, see also Fig.~\ref{fig:fig3}c).
Hence, the gap is assigned to the island with the smaller charging energy.
In order to account for this experimental observation, we resort below to a 
more sophisticated
transport model that includes the ferromagnetic nature of the material.

\begin{table}
\caption{ \label{tab:tab1} Parameters for the small and large Coulomb diamonds 
(CD) extracted from Fig.~\ref{fig:fig3}a)
assuming a double-island structure in the framework of the orthodox theory. The 
charging energy $U=e^2/C_\Sigma$, with $C_{\Sigma} = C_{\rm s} + C_{\rm d} + 
C_{\rm g}$ being the total capacitance, is also given for reference.}
\begin{ruledtabular}
\begin{tabular}{l   l  l}
 \toprule
% \multicolumn{3}{c}{~} \\
\qquad& small CD (ABCD) & large CD (EFGH)\\
% \hline
\midrule
 $C_{\rm d}$  & $5.6 \times 10^{-18}\mathrm{F}$ &  $3.0 \times 
10^{-18}\mathrm{F}$ \\
\midrule
$C_{\rm s}$ 	& $8.4 \times 10^{-18}\mathrm{F}$&   $4.2\times 
10^{-18}\mathrm{F}$ \\
\midrule
$C_{\rm g}$ 	& $28 \times 10^{-20}\mathrm{F}$&   $9 \times 
10^{-20}\mathrm{F}$ \\
\midrule
$U$ 	& $11.2 \times 10^{-3}\mathrm{eV}$&   $21.9\times 10^{-3}\mathrm{eV}$ \\
\bottomrule
\end{tabular}
\end{ruledtabular}
\end{table}

\section{Theoretical modeling}
\label{sect:theoretical modeling}

\begin{figure}[h!]
 \includegraphics[width= \columnwidth]{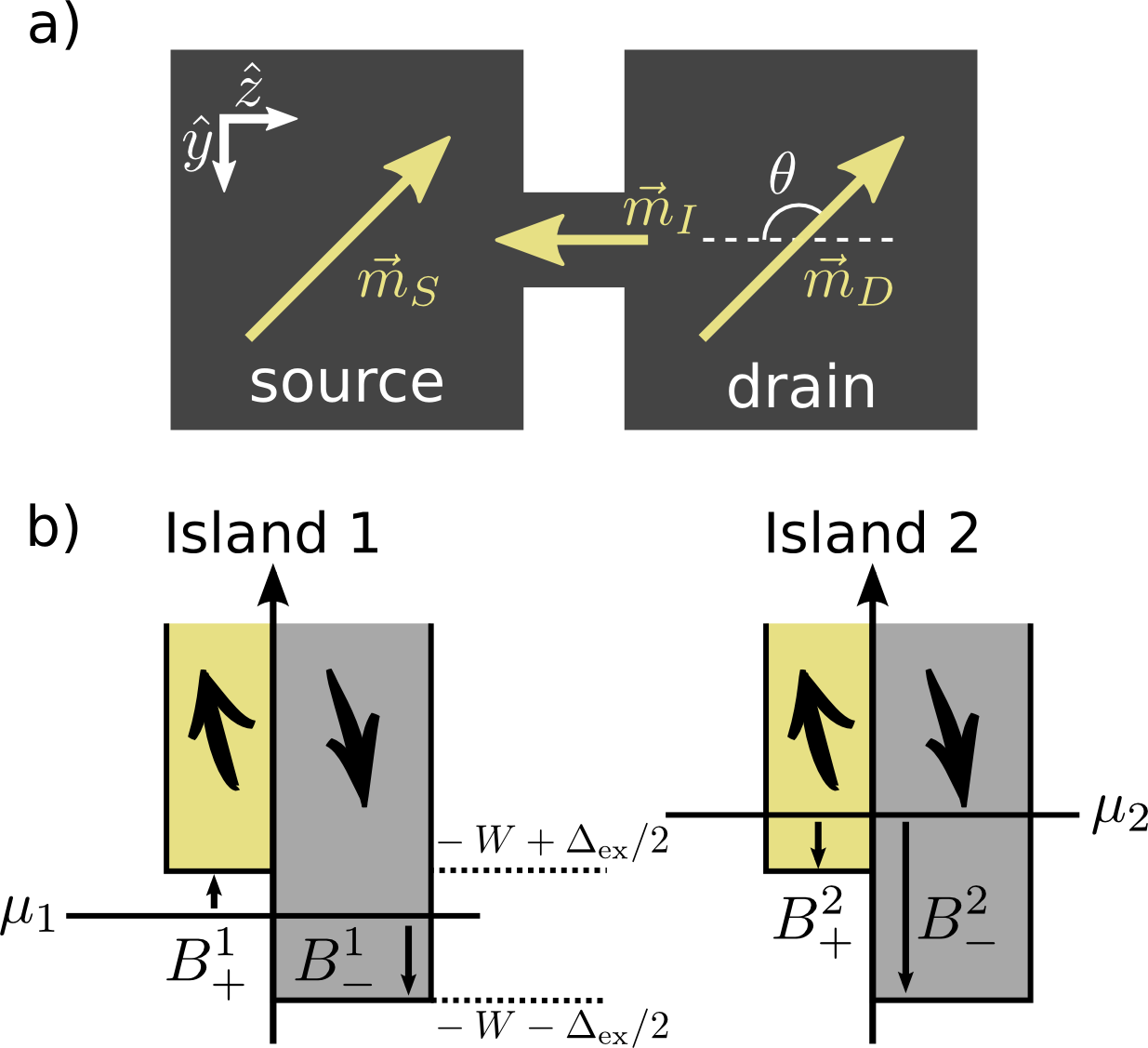}
 \caption{ a) Sketch of the magnetization direction of the leads 
($\vec{m}_{S/D}$) and of the islands ($\vec{m}_I$). The magnetization 
of the leads is determined by the direction of the external magnetic field. 
In the constriction, on the other hand, strain effects are dominating and the 
magnetization direction lies parallel to the constriction axis. In our 
experiment the angle between the two magnetizations  is approximately 
$\theta= \frac{3}{4}\pi$.
 b) Sketch of the density of states of the two metallic islands, with the spins
aligned along the magnetization of the constriction.}
\label{fig:fig5}
\end{figure}

In this section we extend the orthodox theory of Coulomb blockade
\cite{Averin1986, Averin1991,Grabert1991, Grabert1992, Kouwenhoven1997} in 
order to account for the ferromagnetic properties
of the (Ga,Mn)As samples. Although transport through magnetic islands has been 
addressed in the literature,
\cite{Barnas2008} scarce consideration has been given, to our knowledge, to 
the role played by an energy dependent
density of states in the metallic islands. The latter, instead, is crucial to 
explain
the anomalous current blocking observed in the present experiment.

To this end we assume that both leads and the metallic
islands are spin polarized.
Fig.~\ref{fig:fig5}a) shows a sketch of the magnetization 
directions expected in the experiments. 
The magnetization of the ferromagnetic (Ga,Mn)As leads is rather weak, and  can be
tuned by an external magnetic field.  It forms in our experiment an angle of
$45^\circ$ (easy direction) with the transport direction, set by the
longitudinal axis of the NC
($z$-axis, cf. Fig.~\ref{fig:fig5}a)). In the constriction, however, the spin
polarization axis is strongly influenced by strain effects and is expected to be
along the NC longitudinal axis.

In order to explain the blockade effects we claim that 
the angle $\theta$ between the leads and the constrictions magnetization lies in the 
range $\frac{1}{2} \pi < \theta < \frac{3}{2}\pi$.
In other words current suppression originates from the fact, that the majority spin 
carriers in the islands and in the leads have effectively the opposite polarization.
Since only one of
the two superimposed Coulomb diamond structures shows a noteworthy blockade
effect, we conclude, within our model, that the structure with the blockade
stems from transport through a  fully polarized island, while the second
island is only partially polarized.

We describe the islands  polarization with an upward shift in energy of the minority
spin band  with respect to the majority spin band, see Fig.~\ref{fig:fig5}b).
The electro-chemical potential is the external parameter which determines
whether the island is partially or fully polarized.  Partial
polarization is obtained if the chemical potential $\mu_\alpha$ \mbox{($\alpha=1,2$)}
lies
above the bottom of the 
minority
spin band, full polarization when the chemical potential lies between
the bottom of the majority and  of the minority spin bands.

In our model the tunneling of a source electron of the majority spin species
(conventionally the spin up)  to a fully down polarized island is highly
suppressed for low bias voltages since no spin up states are available near
the Fermi level. For bias voltages which are large enough to access
also the
minority spin band ($\alpha_{S} eV_b > B^1_+$, cf. Fig.~\ref{fig:fig5}b)), the
suppression is lifted and an increase of the current is
expected.  For the partially polarized island both spin species can be accessed
already at the Fermi energy and no suppression is observed.

\subsection{Model Hamiltonian}

We describe the nanoconstriction with a system-bath model aimed at mimicking the
structure of the two islands contacted to source and drain leads sketched in 
Fig.~\ref{fig:fig4}a).
The total Hamiltonian is
\begin{equation}
 \hat H = \hat H_{\rm S} + \hat H_{\rm T} + \hat H_{\rm L},
\end{equation}
where
\begin{equation}\label{eq:Hbath}
 \hat H_{\rm L} = \sum_{\eta \in {\rm\{S,D\}} } \sum_{k \sigma} E_{\eta k 
\sigma}
\cdag{\eta k \sigma} \can{\eta k \sigma}
\end{equation}
denotes the Hamiltonian of the two spin polarized leads. We assume to
have a flat, but spin dependent, density of states (\mbox{$\sigma = \up/\down$})
\begin{equation}
\begin{split}
 D_{\eta \up} = \frac{1+p_\eta}{2} D_\eta, \quad
 D_{\eta \down} = \frac{1-p_\eta}{2} D_\eta ,
\end{split}
\end{equation}
which depends on the polarization $p_\eta$ of the leads ($-1\leq p_\eta \leq
1$).
The metallic islands ($\alpha \in \{1,2\}$) in the nanoconstriction
are modeled by
\begin{equation}\label{eq:Hsys}
\begin{split}
 \hat H_{\rm S} =  \sum_{\alpha \in \{1,2\}}& \bigg\{ \sum_{i\tau} 
\epsilon_{\alpha i\tau} \dddag{\alpha i\tau}\dan{\alpha i\tau}  + \alpha_{\rm g}
eV_{\rm g}
\hat N_\alpha  \\
& +
\frac{U_{\alpha}}{2} \hat N_\alpha \big( \hat N_\alpha -1 \big)\bigg\},
\end{split}
\end{equation}
and have in general a different spin quantization axis as the contacts.
We define $\tau = \pm 1$ for spin \makebox{$+/-$},
respectively, using  the  spin-quantization axis of the nanoconstriction.
As already mentioned, we account for the ferromagnetic properties of the 
metallic islands
by assigning spin dependent energy
levels, $\epsilon_{\alpha i \tau}$, and consequently a relative shift of  the
density of states for the two spin directions, $\Deltast$ (Fig.~\ref{fig:fig5}b)).
The long range
Coulomb  interactions are included within a constant interaction model, where
$U_\alpha$ is the charging energy of the island $\alpha$.
The effective coupling of the gate electrode to the metallic islands is taken
into account by the term  proportional to $\alpha_{\rm g} eV_{\rm g}$, with 
$\alpha_{\rm g}=
C_{\rm g}/C_\Sigma$ being
an effective
gate coupling parameter and $V_{\rm g}$  the gate voltage.
The two metallic islands and the leads are weakly coupled by the tunneling
Hamiltonian
\begin{equation}
\begin{split}
\hat H_{\rm T} = \sum_{i \alpha \tau}\sum_{\eta k \sigma} \big(
 t_{\eta \alpha \sigma} u_{\sigma \tau}(\theta)
\cdag{\eta k \sigma} \dan{\alpha i \tau} + \, h.c.
%  +t_{\eta \alpha \sigma} ^* u_{\sigma \tau}^*(\theta) \dddag{\alpha i \tau} \can{\eta
% k \sigma} 
\big),
\end{split}
\end{equation}
where we defined  the function 
$u_{\up +}(\theta) = u_{\down -}(\theta)=\cos(\theta/2) $, $u_{\up -}(\theta) =
u_{\down +}(\theta)= i\sin(\theta/2) $. 
It results from the  non-collinear spin quantization axis of the
islands and the leads. Since the two axis are rotated by an angle of $\theta$ in the
$y$-$z$-plane with respect to each other, the transformation conserves the spin
during tunneling.

\subsection{Density of states of the metallic islands}

Some of the experimental observations can only be understood
if the energy dependence of the density of states, in particular the presence
of different band edges for minority and majority spins, is accounted for.
Specifically, we define the spin-dependent density of
states of island $\alpha$ as:
\begin{equation}\label{eq:dos}
\begin{split}
&g_{\alpha\tau}(\epsilon) =\\
&=
 \tilde g_{\alpha \tau} ~
\Theta\big( \epsilon +W - \tau \Deltast/2\big)
\Theta\big(W + \tau \Deltast/2    - \epsilon \big) \\
&\approx \tilde g_{\alpha \tau} ~
f^-\big(\epsilon + W - \tau \Deltast/2 \big),
\end{split}
\end{equation}
where $W$ is the spin independent contribution to the bandwidth, and $\Deltast$
the exchange band splitting of the ferromagnetic metallic island. The
parameter $\tilde g_{\alpha \tau}$ defines the strength of
the density of sates. 
Since the  $W$ is  the largest energy scale considered in the
following, the upper limit of the density of states can be set to infinity. In
the last line of Eq.~(\ref{eq:dos}) we  have approximated the left Heaviside
function by  $f^- = 1-f^+$, with $f^+$ the  Fermi function; this
allows us to further proceed analytically in the calculation of the transport
properties.
The density of states is also sketched for clarity in Fig.~\ref{fig:fig5}b).
For later reference we define $B_{\tau}^\alpha$ as the energy difference 
between
the bottom of the band of the corresponding spin species $\tau$ and the 
chemical
potential
of the island $\alpha$: $B_{\tau}^\alpha = -W + \tau\Deltast/2 - \mu_\alpha 
$.

\subsection{Transport theory}

In the following we briefly outline the main steps leading to the
evaluation  of the transport characteristics, emphasizing the new ingredients
entering our transport theory. For more details we refer to the Appendix
\ref{app:theory}.
The framework is the orthodox theory of Coulomb blockade
\cite{Averin1986, Averin1991,Grabert1991, Grabert1992, Kouwenhoven1997},
extended to the case of ferromagnetic contacts \cite{Barnas2008} and valid also 
for fully spin polarized metallic islands. The explicit derivation of
the tunnelling rates should illustrate the crucial role
played in our theory by the energy dependent density of states.

The theory is based on a master equation for the reduced density matrix of the 
islands, up to second order in the tunneling Hamiltonian. Since the two metallic
islands are assumed not to interact
with each other, the corresponding density matrices obey independent 
equations of motion (see Appendix \ref{app:theory}).
Moreover, the metallic islands are assumed large enough
to posses a quasi continuous single-particle spectrum, but small enough that
their charging energy dominates the tunnelling processes that change their 
particle number.
We further assume that, in between two tunnelling events, the islands relax to 
a local thermal equilibrium.
Under these assumptions the reduced density matrix of island $\alpha$ can be 
written as
\begin{equation}\label{eq:rho_red_canonical_0}
\begin{split}
\hat \rho^\alpha_{\rm red}(t) = \sum_{N_\alpha} \bigg\{
\mathcal{P}_{N_{\alpha}} \frac{e^{-\beta \hat H_{\rm S,\alpha}} }{
\mathcal{Z}_{N_\alpha}}   \bigg\} P_{N_\alpha}(t),
\end{split}
\end{equation}
where $H_{\rm S,\alpha}$ is the part of the system Hamiltonian associated to the
island $\alpha$, $\mathcal{P}_{N_{\alpha}}$ is the projection operator on the
$N_{\alpha}$-particle subspace and  $\mathcal{Z}_{N_\alpha} = {\rm Tr}_{\rm 
S}\big(
\mathcal{P}_{N_\alpha} e^{-\beta \hat H_{\rm S,\alpha}}\big)$ is the
corresponding (canonical) partition function.
By projecting the master equation on the $N_\alpha$-particle subspace
and tracing over the islands degrees of freedom, we
keep only the occupation probabilities $P_{N_\alpha}$ of finding the island 
occupied by $N_\alpha$ electrons as
dynamical variables. In the stationary limit we find (see 
Appendix~\ref{app:theory})
\begin{equation}\label{eq:master_equation_final}
\begin{split}
\tr{\rm S} \big\{ & \mathcal{P}_{N_{\alpha}}   \dot{\hat\rho}^\alpha_{\infty}
\big\}
 =0 \\
 =  \sum_{ \eta \sigma } &
 \bigg\{
-
\Gamma^{N_{\alpha} \to N_{\alpha}-1  }_{\eta\alpha \sigma}~
P_{N_{\alpha}}
-
  \Gamma^{N_{\alpha} \to N_{\alpha} + 1  }_{\eta\alpha \sigma}~
P_{N_{\alpha}}  \\
& +
\Gamma^{N_{\alpha}-1 \to N_\alpha  }_{\eta\alpha \sigma}~
\, P_{N_{\alpha}-1}
+
 \Gamma^{N_{\alpha} +1 \to N_\alpha }_{\eta\alpha \sigma}~
\,  P_{N_{\alpha} +1}
 \bigg\} .
\end{split}
\end{equation}
Eventually, the stationary current through lead $\eta$ reads
\begin{equation}\label{eq:current_final}
\begin{split}
 I_\eta =&
- e \sum_{\alpha \sigma} \sum_{N_\alpha } \bigg\{  \Gamma^{N_\alpha\to N_\alpha
+1 }_{\eta\alpha \sigma} - \Gamma^{N_\alpha  \to N_\alpha -1
 }_{\eta\alpha \sigma} \bigg\} P_{N_\alpha}.
%  = I_\eta^A +  I_\eta^B,
\end{split}
\end{equation}
In Eqs.~(\ref{eq:master_equation_final}) and (\ref{eq:current_final}) the rates
are defined as
\begin{equation}\label{eq:rates_R}
\begin{split}
&\Gamma^{N_{\alpha} + 1 \to N_{\alpha}}_{\eta\alpha \sigma}
=
 \sum_{\tau }
\frac{1+ \sigma \,p_\eta }{2e^2 R^{\eta \sigma}_{\alpha
 \tau}} ~ |u_{\sigma\tau}(\theta)|^2 ~ 
  b^-( \Delta E_{N_{\alpha}}^G   - \alpha_\eta eV_{\rm b}    ) \\
 &\qquad\times \bigg\{
F(    \Delta E_{N_{\alpha}}^G    + B_\tau^\alpha   - \alpha_\eta eV_{\rm b}  )
-F(   B_\tau^\alpha     )
\bigg\} ,\\
 & \Gamma^{N_{\alpha} \to N_{\alpha} + 1  }_{\eta\alpha \sigma} =
 \sum_{\tau}
 \frac{1+ \sigma \,p_\eta }{2 e^2 R^{\eta \sigma }_{\alpha \tau}}  ~|u_{\sigma \tau
}(\theta)|^2 ~ 
b^+( \Delta E_{N_{\alpha}}^G   - \alpha_\eta eV_{\rm b}    ) \\
&\qquad\times \bigg\{
F(   B_\tau^\alpha    )
-F(    \Delta E_{N_{\alpha}}^G    + B_\tau^\alpha   - \alpha_\eta eV_{\rm b}  )
\bigg\},
\end{split}
\end{equation}
and are expressed in terms of the normal state resistance
$ R^{\eta\sigma }_{\alpha \tau} = \hbar/(2\pi e^2  |t_{ \eta \alpha \sigma}|^2
\tilde g_{\alpha \tau}  D_{\eta})$ and the functions
$b^\pm(x) =  1/(e^{\pm \beta x } -1)$ and $F(x) = x/(e^{\beta x} -1)$,
with $\beta=1/(k_{\rm B}T)$ the inverse temperature. 
We account for the asymmetric bias drop with the bias coupling constants
defined as $\alpha_{\rm {S/D}} = \pm \frac{C_{\rm{d/s}}  + C_{\rm 
g}/2}{C_\Sigma}$.
Further, we defined the grand canonical addition energy
\begin{equation}
\begin{split}
&\Delta E^G_{N_\alpha} =  \alpha_{\rm g} eV_{\rm g}   + U_\alpha N_\alpha + 
\mu_{\alpha} - \mu_0
\\
&= \big(E_{N_\alpha +  1  } - \mu_0  (N_\alpha+1) \big) -  \big( E_{N_\alpha }
-\mu_0 N_\alpha \big)
\end{split}
\end{equation}
which must be paid in order to increase the electron number on island $\alpha$
from $N_\alpha \to N_{\alpha}+1$. We denote $\mu_0$ the chemical potential of 
the
leads at bias $V_{\rm b}=0$.

The rates given in Eq. \eqref{eq:rates_R} differ from the ones of the
orthodox theory of Coulomb Blockade \cite{Averin1986, Averin1991,Grabert1991, 
Grabert1992, Kouwenhoven1997}
even in their spin dependent variation \cite{Barnas2008} due to the energy 
dependent density of states and the explicit dependence on the band edges. The
latter  introduce a new source of current suppression associated to
the absence of states with a specific spin species.
These rates represent the main theoretical contribution of the present work.
For the chemical potential lying far above the bottom of the bands, the
theory recovers again the limit of the classical orthodox theory of Coulomb
blockade. Namely, in the
limit $B \to -\infty$:
\begin{equation}
\lim_{B\to-\infty} \pm b^\pm(x) \big \{F( B ) - F(x+ B) \big\} = F(\pm x).
\end{equation}

\section{Theoretical Results}
\subsection{Comparison with the experiments}
\label{sect:comparison}

\begin{figure}[h!]
 \includegraphics[width= 0.7\columnwidth]{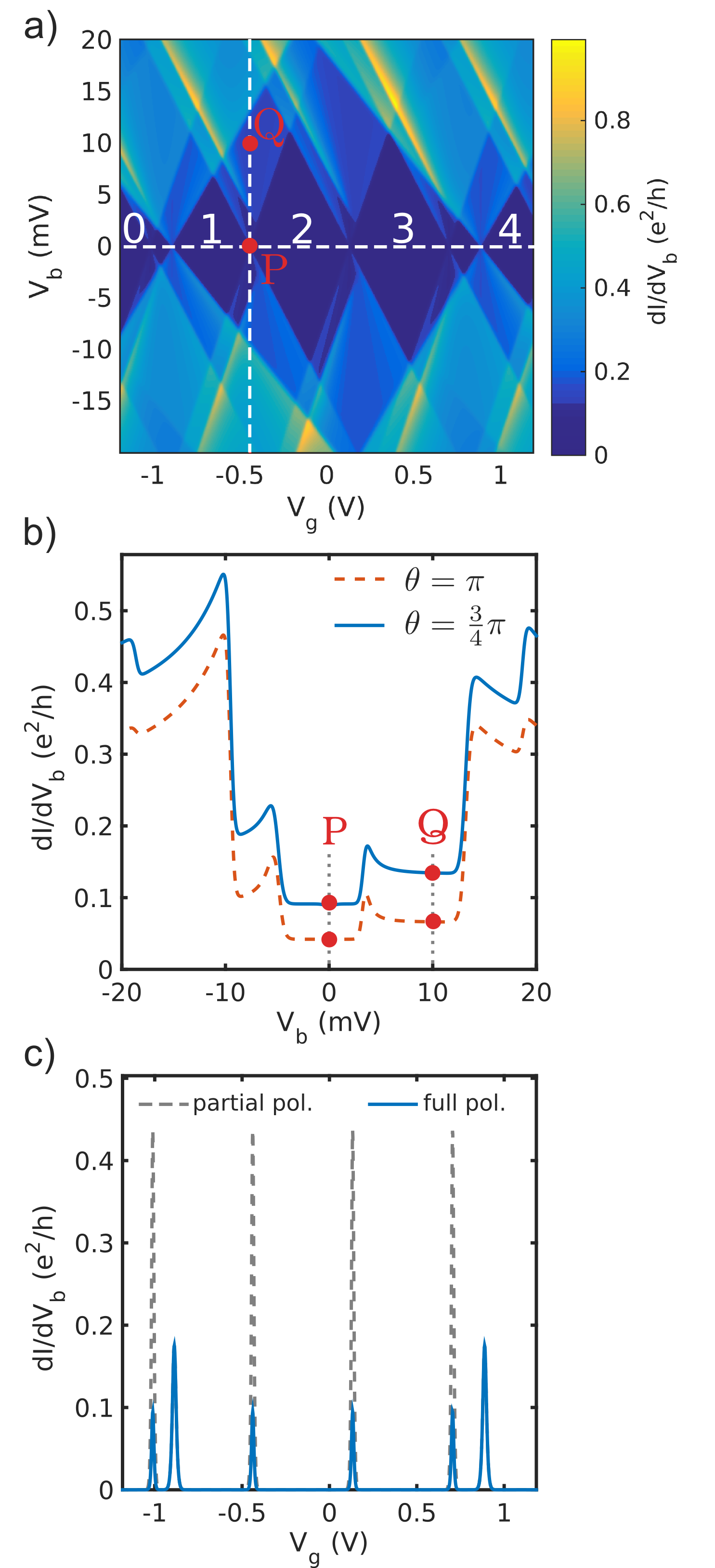}
 \caption{a) Calculated differential conductance of two spin polarized metallic
islands with spin polarized leads. The island with the larger charging energy is
assumed to be partially polarized, while the island with the smaller charging 
energy
is fully polarized. b) Bias trace through a charge degeneracy point of the fully
polarized island. The gate position of the line trace is marked as a dashed
line in a).
c) Gate traces at $V_b =0$ for a full (solid blue line) and a partial
polarization  (dashed grey line) of island $1$. Island $2$ remains partially
polarized. In the fully polarized cased the conductance peaks of island $1$  are
suppressed with respect to the partially polarized case. For island $2$ both curves
are identical. 
The parameters used to obtain this
figure are: $\alpha_{{\rm S}1} = 0.4$, $\alpha_{{\rm S}2} = 0.42$, $U_1 = 11.2$ 
meV
and
$U_2 = 21.9$ meV in accordance with the parameters for the capacitive couplings 
of
Table \ref{tab:tab1}. Moreover $R_{1\tau}^{\eta \sigma} = 0.57 \cdot10^3 h/e^2$,
$R_{2\tau}^{\eta\sigma} = 1.4\cdot10^3 h/e^2$, $B^1_+= 2 $\,meV,
$B^1_- 
= -18 $\,meV,
$B^2_+ = -10 $\,meV, $B^2_- = -35 $\,meV, $\mu_1 = -42$\,meV, $\mu_2 
= -32$\,meV,
$p_\eta=0.8$, and $k_{\rm B}T = 0.07$\,meV.
For the full polarized island $1$ in c) $B^1_+ = -10 $\,meV, $B^1_- 
= -30 $\,meV. 
}
\label{fig:fig6}
\end{figure}

The results of our simulation are reported in Fig.~\ref{fig:fig6}a), with the
differential conductance shown as a function of the bias and gate voltage.
We see the same nested diamond structure as in the experiments. In our theory
the diamonds at the charge degeneracy points labeled 0-1 and 3-4 close. 
Between the diamonds 1-2 and 2-3 the differential conductance is suppressed 
for bias voltages smaller than a certain threshold bias.
 Fig.~\ref{fig:fig6}b) shows a bias trace calculated at the charge degeneracy 
point 1-2, for two different angles $\theta$ between the  magnetization 
vectors of the leads, $\vec{m}_\alpha$,  and the metallic islands, $\vec{m}_I$. 
 It shows a suppression of the differential conductance at 
point (P) with respect to point (Q). The width of the suppression region 
corresponds to the one observed experimentally in Fig.~\ref{fig:fig3}b)
 and is proportional to $B^1_+$, 
 the energy difference between the bottom of the minority 
spin band and the chemical potential of island $1$ (cf.
Fig.~\ref{fig:fig5}b)). 
In contrast to the experiments no full blockade can be observed at (P).
A change of the orientation of the magnetization directions from 
$\theta= \pi $ (dashed red line) to $\theta= \frac{3}{4} \pi$ (solid blue line) is 
shifting the curve upwards. Besides the constant shift the two curves are 
qualitatively the same. 

To emphasize the effect of the islands degree of polarization on the 
suppression
mechanism, a conductance trace at $V_b=0$ of a full  polarized 
island $1$ is compared to the case of a  partial polarized island $1$ in
Fig.~\ref{fig:fig6}c). Partial polarization is achieved by shifting the 
electrochemical potential of island $1$ by $12$\,meV  up in energy.
The solid blue line shows the full polarized case, where the two larger peaks 
correspond to the larger Coulomb diamond (island $2$). 
The peak observed in the experiment (Fig.~\ref{fig:fig3}c)) 
we ascribe to transport across this partially polarized island. Although the
theoretically predicted second peak is missing in Fig.~\ref{fig:fig3}c) we note that
the corresponding blockade between diamond 3 and 4 is much less pronounced than
between, e.g. 2 and 3. This asymmetry between the degeneracy points 0-1 and 3-4,
however, cannot be accounted for by our model which predicts a periodicity of the
Coulomb oscillation pattern. The four smaller peaks in Fig.~\ref{fig:fig6}c) belong
to the smaller Coulomb diamond structure, corresponding to island $1$, i.e. the fully
polarized one (Fig.~\ref{fig:fig5}b)). Even though the conductance is not completely
suppressed as in the experiment, the conductance peaks are strongly reduced with
respect to the partially polarized case. In the latter
(dashed grey lines) no suppression  is present and the conductance peaks of 
island $1$ are  by a factor of $4$ larger. Below we address a possible reason for
the incomplete blocking within the model.
 Since the parameters of island $2$ are kept the same,  both for the fully and
partially polarized cases the corresponding conductance peaks are not changing. 
Despite the fact that a comparison
of the calculated gate trace to the experimental one  in Fig.~\ref{fig:fig3}c)
reveals some limitations of the model, the essential feature, i.e. the suppression
inside the large Coulomb diamond, is reproduced.

\subsection{The mechanism of current suppression}

\begin{figure}
 \includegraphics[width=0.9\columnwidth]{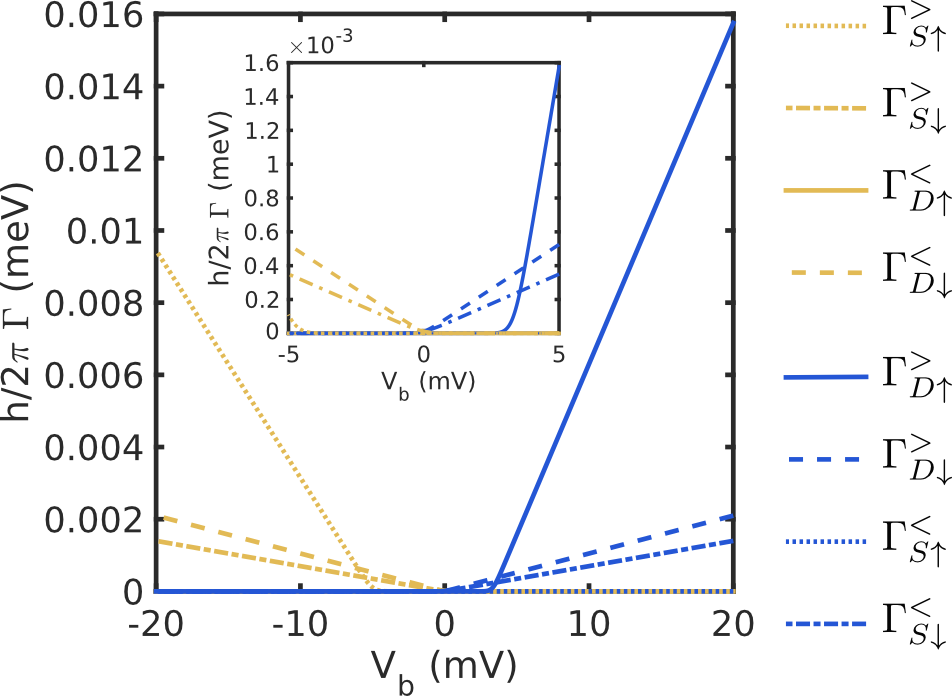}
\caption{ All tunnelling rates for the island $1$ plotted at a charge degeneracy
point as a function of the bias voltage $V_{\rm b}$. The angle between the two
magnetization direction is $\theta=\pi$. }
\label{fig:fig7}
\end{figure}

For a better understanding of the mechanism underlying the blockade,  we
derive analytically the differential conductance for the island $1$  at the two
points
(P) and (Q) marked  in Fig.~\ref{fig:fig6}b).
For simplicity the case $\theta= \pi$ is considered, since  qualitatively the 
blockade mechanism is the same in both cases.

Notice that both P and Q correspond to a gate voltage such that
 $\Delta E^G_{N} = 0$,  {\it i.e.} at the charge degeneracy point of the
\mbox{\mbox{$N$}-\mbox{$N+1$}} transition. To obtain the differential
conductance, according to Eq.~(\ref{eq:master_equation_final}) and
(\ref{eq:current_final}), the transition rates  $ \Gamma^{N
\to N \pm 1}_{\eta \alpha \sigma} $ are required. For simplicity we have dropped 
the
subscript $1$ from the
excitation energy $\Delta E^G_{N}$ since we will refer from now on always to the 
same island.

In Fig.~\ref{fig:fig7} we show the  transition rates as a
function of the bias $V_{\rm b}$. To simplify the notation, we  replaced $ \Gamma^{N
\to N \pm 1}_{\eta \alpha \sigma} \to \Gamma^{\gtrless }_{\eta
\sigma}$.
Notice their linear dependence on the bias above a certain threshold.
Thus, in that bias range one can approximate them as:
\begin{equation}\label{eq:rates_linear_approx}
\begin{split}
&\Gamma_{\rm S\down}^< = -\mc{B}_{\rm S \down} V_{\rm b},\\
&\Gamma_{\rm D\down}^> = \mc{B}_{\rm D \down} V_{\rm b},\\
&\Gamma_{\rm D\up}^> = \mc{A}_{\rm D \up}+ \mc{B}_{\rm D \up} V_{\rm b},\\
\end{split}
\end{equation}
where $\mc{A}_{\rm D\up}$ is a constant accounting for the threshold bias,
and
\begin{equation}\label{eq:def_slope_rates}
 \mc{B}_{\eta \sigma} = \frac{2\pi e}{\hbar^2} D_{0}\frac{1+\sigma p}{2}
\tilde{g}_\sigma \alpha_\eta |t_\eta|^2 .
\end{equation}
Here, $D_0 = D_\eta$ is assumed to be independent of the lead.
For the  point (P) within the first plateau  only the rates with
$\sigma = \down$, namely $\Gamma^>_{\rm D\down}$ and $\Gamma^<_{\rm S\down}$, are
nonzero.
Hence, according to the principle of detailed balance: $\Gamma_{\rm D\down}^> P_N = \Gamma_{\rm S\down}^< P_{N+1}$.
Imposing probability conservation we find $P_N= \Gamma_{\rm S\down}^</(\Gamma_{\rm D\down}^> +\Gamma_{\rm S\down}^<)$.
Thus the stationary current equals $I_D^{(P)} =  -e\Gamma_{\rm D\down}^> P_N =  -e\Gamma_{\rm D\down}^>
\Gamma_{\rm S\down}^</(\Gamma_{\rm D\down}^> +\Gamma_{\rm S\down}^<) \propto 
(1-p)^2$,
which  is suppressed for a large spin polarization $p$. Here the polarization
$p$ is assumed to be equal for both leads.

At the point (Q), only one additional rate, $\Gamma^>_{\rm D\up}$, is contributing (the
rate $\Gamma^<_{S\up}$  is zero due to the lower
bound of the density of states).
In this  bias range the equations of detailed balance and probability
conservation yield
$
 P_N= \Gamma_{\rm S\down}^</(\Gamma_{\rm D\down}^>+ \Gamma_{\rm D\up}^>
+\Gamma_{\rm S\down}^< ).
$
The resulting stationary current is then
$
I_D^{(Q)} = -e( \Gamma_{\rm D\down}^> + \Gamma_{\rm D\up}^> ) P_N =
-e(\Gamma_{\rm D\down}^>  + \Gamma_{\rm D\up}^>)
\Gamma_{\rm S\down}^< /(\Gamma_{\rm D\down}^> + \Gamma_{\rm D\up}^>+\Gamma_{\rm S\down}^<  )
\propto  (1-p).
$
Again the current is suppressed for large spin polarization.

Inserting Eq.~(\ref{eq:rates_linear_approx}) into the current expressions at the
points P and Q
we find
\begin{equation}\label{eq:IP_linear}
I_D^{(P)} =   e\frac{ \mc{B}_{\rm S\down} \mc{B}_{\rm D\down}}{\mc{B}_{\rm D\down}
-\mc{B}_{\rm S\down}  } V_{\rm b}
\end{equation}
and
\begin{equation}\label{eq:IQ_linear}
\begin{split}
& I_D^{(Q)} = e\frac{+\mc{B}_{\rm S\down}\mc{A}_{\rm D\up} V_{\rm b} - \mc{B}_{\rm S\down} (
\mc{B}_{\rm D\up} + \mc{B}_{\rm D\down}  ) V_{\rm b}^2 }{ \mc{A}_{\rm D\up}
+  ( \mc{B}_{\rm D\up} + \mc{B}_{\rm D\down} -\mc{B}_{\rm S\down}) V_{\rm b}}.
\end{split}
\end{equation}

Taking the ratio of the two differential conductance plateaus, {\it i.e.} the ratio
of Eqs.~(\ref{eq:dIP_linear}) and (\ref{eq:dIQ_linear}), we find
\begin{equation}
\begin{split}
& { \frac{dI_D^{(P)} }{dV_{\rm b}} }\big/
\frac{dI_D^{(Q)} }{dV_{\rm b}}  =
\frac{1}{\alpha_D |t_{\rm D}|^2 - \alpha_{\rm S} |t_{\rm S}|^2 }\\
&\times  \bigg(
\alpha_D |t_{\rm D}|^2 -\alpha_{\rm S} |t_{\rm S}|^2  \frac{(1-p) \tilde{g}_\down}{(1+p) \tilde{g}_\up +
(1-p) \tilde{g}_\down}
\bigg).
\end{split}
\end{equation}
Thus, within our simple model, the ratio $\mathcal{R}$ of the height of  the two
plateaus is limited by
\begin{equation}\label{eq:ratio}
\frac{\alpha_D |t_{\rm S}|^2}{\alpha_D |t_{\rm D}|^2 - \alpha_{\rm S} |t_{\rm 
S}|^2 } \leq \mathcal{R} \leq 1.
\end{equation}
In other words the ratio
of the two differential conductance plateaus is limited in our theory, leading
to some discrepancy with the experimentally observed ratio, cf. points (P) and (Q)
marked in Fig.~\ref{fig:fig3}b).
Since the parameters $\alpha_\eta$ are determined experimentally, the only
possibility to change the ratio is to modify the coupling constants $|t_\eta|$.
However, the increase of the  coupling constants necessary to fit the
experimental value, would lead to a huge asymmetry in the stability diagram which
is not observed experimentally. Despite the  discrepancy
between $\mathcal{R}$ and the experimental ratio, we think that the theory 
clearly suggests a
mechanism which can lead to a suppression of the conductance due to spin
polarization in the framework of an orthodox theory of Coulomb blockade.
To better fit the experiments a more realistic energy
dependence of the density of states which also accounts for valence bands is necessary. With such an
energy dependence the rates can change their slope as a function of the bias
voltage, leading to an even more pronounced bias dependent suppression of the differential
conductance.

\section{Conclusion}
\label{sect:conclusions}

In this work we have reported on a detailed study of the transport characteristics of
nanofabricated narrow constrictions in (Ga,Mn)As thin films. By means of a two step
electron beam lithography technique we have fabricated well
defined nanoconstrictions  of different sizes. Depending
on channel width and length, for a specific material, different low-temperature
transport regimes have been identified, namely the ohmic regime, the
single electron tunnelling regime (SET) and a completely insulating regime.
In the SET, complex stability diagrams with nested Coulomb diamonds and anomalous
conductance suppression in the vicinity of charge degeneracy points have been measured.

In order to rationalize these observations we proposed, for a specific
nanoconstriction, a model consisting of two ferromagnetic islands coupled to
ferromagnetic leads. 
In particular, the  angle $\theta$ between the leads and the islands magnetization
lies in the range $ \frac{1}{2}\pi < \theta < \frac{3}{2} \pi$. 
Moreover, the full polarization of one of the metallic islands is crucial.
We studied the transport characteristics of the system in terms of a modified
orthodox theory of Coulomb blockade
which takes into account the energy dependence of the density of states in the
metallic islands. The latter represents an important generalization of existing formulations and
is determinant for the qualitative understanding of the present experiments. In fact,
the explicit appearance of the minority
spin band edge in the expression of the tunnelling rates yields a pronounced
conductance suppression at the charge degeneracy points.
To account for the full suppression of conductance observed in the experiments the
simple model used in this work should  be further improved. For example
the hole character of the charge carriers and associated spin orbit coupling
effects are not captured by our  model. 
Furthermore, it is straightforward  to  combine the present theory  with
microscopic models that allow for a realistic description of the islands density of
states.

We acknowledge for this work the financial support of the Deutsche Forschungsgemeinschaft under the
research programs SFB 631 and SFB 689.

\appendix

\section{Sample fabrication}
\label{app:sample_fab}
\subsection{Two-step EBL fabrication process}
Both steps are based on the standard EBL resist poly-methyl-methacrylate (PMMA).
In the first step one exposes the resist using an extremely high line-dose
(approx. $30.000 \text{pC/cm}$) in order to define a narrow crosslinked
PMMA-line. This line is very robust and does not get  removed by common
organic solvents
like acetone. Hence, after cleaning the sample in a bath of acetone, the
crosslinked PMMA-line remains on top of the sample while the unexposed PMMA is
removed from the sample surface. For the second step the sample is again coated
with a fresh layer of PMMA resist. This time one uses a common dose (approx.
$2000 \text{pC/cm}$) in order to expose a second line perpendicular to the
crosslinked one. After removing the exposed resist using a standard developer
solution consisting of isopropyl alcohol and Methyl-isobutyl-ketone (MIBK), we
get the patterned mask for the subsequent ion-beam-etching, shown in
Fig.~\ref{fig:fig1}a).

\section{Equation of motion for a orthodox theory of Coulomb blockade}
\label{app:theory}

In this appendix we derive an extension of the orthodox theory of Coulomb
blockade for the case of spin polarized contacts as well as of a spin polarized
metallic island. In particular we will consider explicitly the lower bound of
the density of states in the metallic island.

The transport theory is based on the Liouville-von Neumann equation for the
reduced density matrix in the interaction picture
\begin{equation}\label{eq:Liouville}
 i\hbar \frac{\partial}{\partial t} \hat \rho_{\rm I}(t) =  \big  [ \hat
H_{\rm T,I}(t),
\hat \rho_{\rm I}(t) \big],
\end{equation}
which we expand to second order in the tunneling Hamiltonian $\hat H_{\rm T}$.
Prior to $t=0$ the system and the leads do not interact and the density matrix
can be written as a tensor product of the density matrices of the subsystems
\begin{equation}
 \hat \rho = \hat \rho_S(0) \otimes \hat \rho_{\rm L} \equiv \hat \rho_{\rm S}(0)
\hat \rho_{\rm L}
\end{equation}
Since the  leads are considered thermal baths of noninteracting fermions,
$\hat \rho_{\rm L} $ reads
\begin{equation}
 \hat \rho_{\rm L} = \frac{e^{-\beta( \hat H_{\rm L}- \sum_\eta\mu_\eta \hat N_\eta)
}}{\mathcal{Z}_{\rm L,G}}.
\end{equation}
Further, we assume that due to fast relaxation processes in the leads, the
density matrix can be written as
$\hat \rho_{\rm I}(t) =  \hat \rho_{\rm red,I}(t) \hat \rho_{\rm L} + \mathcal{O}(\hat H_{\rm T})
$
, with $\hat{\rho}_{\rm red,I} = {\rm Tr}_{\rm L} \hat{\rho}$. Moreover, due to the independence of the two metallic
islands $\hat \rho_{\rm red}(t) = \hat \rho_{\rm red}^1(t) \hat \rho_{\rm red}^2(t)$ and each component obeys the following equation of motion:
\begin{equation}\label{eq:master_equation_0}
\begin{split}
&\dot{\hat\rho}^\alpha_{\rm red}(t) = -\frac{i}{\hbar} \big[ \hat H_{\rm S},
\hat
\rho^\alpha_{\rm red}(t)
\big] \\
&-\frac{1}{\hbar^2} \int_0^t dt'' \tr{\rm L} \bigg\{ \bigg[ \hat H_{\rm T} ,\big[
H_{\rm T,I}(-t''), \hat \rho^\alpha_{\rm red}(t) \rho_L   \big] \bigg] \bigg\},
\end{split}
\end{equation}
where $\alpha = 1,2$ labels the metallic island.

For the system we assume that the metallic islands are  large enough
to posses a quasi continuous single-particle spectrum, but small enough that
their charging energy dominates the tunnelling processes that change their 
particle number.
Furthermore, it is assumed that the islands will relax to a local thermal 
equilibrium on a time scale shorter than the inverse of the average electronic 
tunnelling rate.
Under these assumptions, the reduced density matrix
can be written as
\begin{equation}\label{eq:rho_red_canonical2}
\begin{split}
\hat \rho^\alpha_{\rm red}(t) =    \sum_{N_\alpha}
\mathcal{P}_{N_{\alpha}} \frac{e^{-\beta \hat H_{{\rm S},\alpha }}
}{\mathcal{Z}_{N_\alpha}}
P_{N_\alpha}(t) ,
\end{split}
\end{equation}
with $\mathcal{Z}_{N_\alpha}  =\tr{\rm S} \big\{\mathcal{P}_{N_{\alpha}}
e^{-\beta \hat H_{\mathrm{S},\alpha}}\big\}  $, and
\begin{equation}
  \mathcal{P}_{N_\alpha }=  \sum_{ \substack{ \{n_i\}_\alpha \\\sum_i n_i =
N_\alpha
 } } \ketbra{ \{n_i \}_\alpha } ,
\end{equation}
is the projection operator on the $N_\alpha$-particle subspace.
Notice that in Eq.~\eqref{eq:rho_red_canonical2}, due to the projector operator $\mathcal{P}_{N_\alpha}$, the
only statistically relevant term of the system Hamiltonian 
$\hat{H}_{\mathrm{S},\alpha  }$ is $\hat h_{\rm S}^\alpha =
\sum_{i\sigma}\epsilon_{\alpha i \sigma} d^{\dagger}_{\alpha i \sigma} d_{\alpha i \sigma}$.
The term $
e^{-\beta(\frac{U_\alpha}{2}N_\alpha(N_\alpha -1) + \alpha_{\rm g} eV_{\rm g}
N_\alpha)}   $ becomes a constant and is canceling out in the density matrix.
Inserting explicitly $\hat H_{\rm T}$ in Eq.~\eqref{eq:master_equation_0}, we find

\begin{widetext}
~
\begin{equation}\label{eq:master_equation_5}
\begin{split}
\tr{\rm S} \big\{ \mathcal{P}_{N_{\alpha}}  \dot{\hat\rho}^\alpha_{\rm red}(t)  \big\}
=&
-\frac{1}{\hbar^2}  \sum_{ \eta \eta'}\sum_{{ \substack{ k i
\sigma \tau \\ k' i' \sigma' \tau'}  } }   t_{\eta \alpha \sigma} u_{\sigma
\tau}(\theta) ~t_{\eta' \alpha \sigma'}^* u_{\sigma'\tau'}^*(\theta)
\int_0^t dt''  \bigg\{ \\ &
\tr{\rm S} \bigg\{ \mathcal{P}_{N_{\alpha}}
\dddag{\alpha i \tau} \dan{\alpha i'\tau',I}(-t'')
\hat\rho^\alpha_{\rm red}(t)  \bigg\}
\tr{\rm L} \big\{
\can{\eta k \sigma} \cdag{\eta'k'\sigma',{\rm I}}(-t'') \hat\rho_{\rm L}  \big\} \\
+&
\tr{\rm S} \bigg\{ \mathcal{P}_{N_{\alpha}}
 \dan{\alpha i \tau} \dddag{\alpha i'\tau',{\rm I}}(-t'')
 \hat \rho^\alpha_{\rm red}(t) \bigg\}
\tr{\rm L} \big\{
\cdag{\eta k \sigma} \can{\eta'k'\sigma',{\rm I}}(-t'')\hat\rho_{\rm L} \big\}  \\
-&
\tr{\rm S} \bigg\{\dan{\alpha i'\tau',{\rm I}}(-t'')  \mathcal{P}_{N_{\alpha}}
\dddag{\alpha i \tau} \, \rho^\alpha_{\rm red}(t)
\bigg\}
 \tr{\rm L} \big\{ \cdag{\eta'k'\sigma',{\rm I}}(-t'')
\can{\eta k \sigma} \,
\hat\rho_{\rm L} \,
\big\} \\
-&
\tr{\rm S} \bigg\{   \dddag{\alpha i'\tau',{\rm I}}(-t'')  \mathcal{P}_{N_{\alpha}}
\dan{\alpha i \tau}\,
\hat\rho^\alpha_{\rm red}(t)
\bigg\}
 \tr{\rm L} \big\{ \can{\eta'k'\sigma',{\rm I}}(-t'')
\cdag{\eta k \sigma} \,
\hat\rho_{\rm L} \,
 \big\}
\\
+& c.c.
 \bigg\} .
\end{split}
\end{equation}
\end{widetext}
In the following we are analyzing  the first term of
Eq.(\ref{eq:master_equation_5}) in more detail, the other terms can be
evaluated in complete analogy.
The calculation of the trace over the lead degrees of freedom gives
\begin{equation}
\begin{split}
&\tr{\rm L} \big\{
\can{\eta k \sigma} \cdag{\eta'k'\sigma',{\rm I}}(-t'')\hat\rho_{\rm L} \big\} \\
&=
e^{\ihbar E_{\eta k} (-t'') } f^-(E_{\eta k } - \mu_\eta)
\delta_{kk'}\delta_{\eta \eta'} \delta_{\sigma \sigma'},
\end{split}
\end{equation}
where the time evolution of the creation and annihilation operators of
the leads is given by $ \cdag{\eta k \sigma,{\rm I}}(t) = e^{\ihbar E_{\eta k} t } \cdag{\eta k \sigma}$.
For the system operators the time evolution can be carried out in a similar way,
keeping in mind that the parts proportional to the total number operator can
be factorized
\begin{equation}\label{eq:average_sys}
\begin{split}
&\tr{\rm S} \bigg\{ \mathcal{P}_{N_{\alpha}}
\dddag{\alpha i \tau} \dan{\alpha i'\tau',{\rm I}}(-t'')
\hat\rho^\alpha_{\rm red}(t)  \bigg\} \\
&=
e^{\ihbar ( \epsilon_{\alpha i'\sigma'} + \alpha_{\rm g}  eV_{\rm g}  + U (N_{\alpha} -1) )
t'' }
\\
& \times \tr{\rm S} \bigg\{ \mathcal{P}_{N_{\alpha}}
\dddag{\alpha i \tau} \dan{\alpha i'\tau'}
\hat\rho^\alpha_{\rm red}(t)  \bigg\}.
\end{split}
\end{equation}
In order to perform the trace over the system degrees of freedom another
approximation is necessary.
By taking the average in the grand canonical ensemble, the particle number is
determined by the chemical potential and we can remove the projection operator:
\begin{equation}
\begin{split}
&\tr{\rm S} \bigg\{ \mathcal{P}_{N_{\alpha}}
\dddag{\alpha i \sigma} \dan{\alpha i'\sigma',{\rm I}}(-t'')
\hat\rho^\alpha_{\rm red}(t)  \bigg\}\\
&= \tr{\rm S}
\bigg\{ \mathcal{P}_{N_{\alpha}}
\dddag{\alpha i \sigma} \dan{\alpha i'\sigma',{\rm I}}(-t'')
\frac{e^{-\beta\hat h_{\rm S}^\alpha}}{Z_{N_\alpha}}\bigg\}  P_{N_\alpha} \\
&\approx \tr{\rm S}
\bigg\{ \dddag{\alpha i \sigma} \dan{\alpha i'\sigma',{\rm I}}(-t'')
\frac{e^{-\beta(\hat h_{\rm S}^\alpha-\mu_{\alpha,N_\alpha})}}{Z_{\mu_{\alpha, 
N_\alpha}}}\bigg\} P_{N_\alpha} 
\end{split}
\end{equation}
This approximation becomes exact in the limit of  \mbox{$N\to\infty$}. 
In presence of a
quasi-continuous energy spectrum of the islands we can further drop the $N_\alpha$ dependence of the
chemical potential, for small relative variations of $N_\alpha$. 

The trace in Eq.~(\ref{eq:average_sys}) can now be
evaluated in the standard way and it yields Fermi functions.
Inserting the results for the traces in \Eq{\ref{eq:master_equation_5}} we
obtain:

\begin{widetext}
 \begin{equation}\label{eq:master_equation_6}
\begin{split}
\tr{\rm S} \big\{ \mathcal{P}_{N_{\alpha}}  \dot{\hat\rho}^\alpha_{\rm red}(t)  \big\}
=&
-\frac{1}{\hbar^2}  \sum_{ \eta }\sum_{{  k i
\sigma  \tau} }   ~ |t_{\eta \alpha  \sigma}|^2~ |u_{\sigma
\tau}(\theta)|^2
\int_0^t dt''  \bigg\{ \\ &
 e^{\ihbar (-E_{\eta k}+\epsilon_{\alpha  i \tau} + \alpha_{\rm g} eV_{\rm g}  +
U_\alpha (N_{\alpha} -1) )t'' } f^+(\epsilon_{\alpha  i \tau} -\mu_\alpha
)
   f^-(E_{\eta k } - \mu_\eta )\, P_{N_{\alpha}}(t)   \\
+&
e^{ -\ihbar  ( -E_{\eta k } + \epsilon_{\alpha  i \tau} + \alpha_{\rm g} eV_{\rm g} +
U
N_{\alpha} )t'' }  f^-(\epsilon_{\alpha  i \tau} - \mu_\alpha)
 f^+(E_{\eta k} -\mu_\eta )\,  P_{N_{\alpha}}(t)  \\
-&
e^{\ihbar(-E_{\eta k} +\epsilon_{\alpha  i \tau } + \alpha_{\rm g} eV_{\rm g}  +
U(N_{\alpha}-1 ) )
t'' }\,
  \, f^-(\epsilon_{\alpha  i \tau}-\mu_\alpha)
 f^+(E_{\eta k } -\mu_\eta   ) \, P_{N_{\alpha}-1}(t) \\
-&
e^{-\ihbar( -E_{\eta k}   + \epsilon_{\alpha  i \tau}  + \alpha_{\rm g} e V_{\rm g} +
U N_{\alpha}  ) t'' } f^+(\epsilon_{\alpha  i\tau} - \mu_{S\alpha  } )
f^-(E_{\eta k} - \mu_\eta) \,  P_{N_{\alpha} +1}(t)
\\
+& c.c.
 \bigg\} .
\end{split}
\end{equation}

\end{widetext}

Since we are only interested in the stationary solution of the master equation,
we send $t\rightarrow \infty$ and use the Dirac identity
\begin{equation}
\int_0^\infty dt e^{i \omega t }  = \pi \delta(\omega) + i \,\lim_{\eta \to 0}
\mathrm{Im}\bigg( \frac{i}{\omega + i \eta} \bigg)
\end{equation}
to evaluate the integrals.
Due to statistical averages no coherences are possible in the master equation
and the two complex conjugated parts can be summed up. We find
\begin{widetext}
\begin{equation}\label{eq:master_equation_7}
\begin{split}
\tr{\rm S} \big\{ \mathcal{P}_{N_{\alpha}}  \dot{\hat\rho}^\alpha_{\infty} \big\}=0
=&
-\frac{2\pi}{\hbar}  \sum_{ \eta }\sum_{{  k i
\sigma \tau  } }  ~   |t_{\eta \alpha  \sigma}|^2 ~  |u_{\sigma \tau}(\theta)|^2
 \bigg\{ \\ &
 \delta \big(-E_{\eta k} +\epsilon_{\alpha  i \tau} + \alpha_{\rm g} eV_{\rm g}  +
U_\alpha (N_{\alpha} -1) \big) \, f^+(\epsilon_{\alpha  i \tau} -\mu_{\alpha
} )
   f^-(E_{\eta k } - \mu_\eta )\, P_{N_{\alpha}}  \\
+&
\delta  \big( -E_{\eta k } + \epsilon_{\alpha  i \tau} + \alpha_{\rm g} eV_{\rm g} +
U
N_{\alpha} \big) \, f^-(\epsilon_{\alpha  i \tau} - \mu_\alpha)
 f^+(E_{\eta k} -\mu_\eta )\,  P_{N_{\alpha}} \\
-&
\delta\big(-E_{\eta k} +\epsilon_{\alpha  i \tau} + \alpha_{\rm g} eV_{\rm g}  +
U(N_{\alpha}-1 ) \big)
  \, f^-(\epsilon_{\alpha  i \tau}-\mu_\alpha)
 f^+(E_{\eta k } -\mu_\eta   ) \, P_{N_{\alpha}-1} \\
-&
\delta\big( -E_{\eta k}   + \epsilon_{\alpha  i \tau}  + \alpha_{\rm g} e V_{\rm g} +
U N_{\alpha}  \big) \,f^+(\epsilon_{\alpha  i\tau} - \mu_{\alpha  } )
f^-(E_{\eta k} - \mu_\eta) \,  P_{N_{\alpha} +1}
 \bigg\}.
\end{split}
\end{equation}
\end{widetext}
Further, we consider the continuum limit of the states in the quantum dot
\begin{equation}
\sum_i \to \int_{-\infty}^\infty d\epsilon
~g_{\alpha \tau}(\epsilon ),
\end{equation}
with $g_{\alpha \tau} (\epsilon) $ being the energy dependent density of
states in island $\alpha $ with the spin $\tau$, defined in Eq.~(\ref{eq:dos}). For
the leads
\begin{equation}
\sum_k \to \int_{-\infty}^{\infty} d E ~D_{\eta \sigma},
\end{equation}
where $D_{\eta \sigma}$ is the density of states of lead $\eta$ which is
considered in the flat band limit.
The integration over the lead degrees of freedom gives:
\begin{widetext}

\begin{equation}\label{eq:master_equation_9}
\begin{split}
\tr{\rm S} \big\{ \mathcal{P}_{N_{\alpha}}  \dot{\hat\rho}^\alpha_{\infty}  \big\}
=0=&
-\frac{2\pi}{\hbar}  \sum_{ \eta \sigma \tau } ~|t_{\eta
\alpha \sigma}|^2 ~ |u_{\sigma\tau}(\theta)|^2  ~ D_{\eta
\sigma}
\int d\epsilon ~
g_{\alpha \tau}(\epsilon)
 \bigg\{ \\ &
f^+(\epsilon -\mu_{\alpha } )
   f^-(\epsilon + \Delta E_{N_{\alpha}-1} - \mu_\eta )\,
P_{N_{\alpha}}  \\
+&
f^-(\epsilon - \mu_{\alpha })
 f^+(\epsilon + \Delta E_{N_{\alpha}} -\mu_\eta )\,  P_{N_{\alpha}}  \\
-&
 f^-(\epsilon-\mu_{\alpha })
 f^+(\epsilon + \Delta E_{N_{\alpha}-1} -\mu_\eta   ) \, P_{N_{\alpha}-1} \\
-&
f^+(\epsilon - \mu_{\alpha  } )
f^-(\epsilon  + \Delta E_{N_{\alpha}}  - \mu_\eta) \,  P_{N_{\alpha} +1}
 \bigg\} .
\end{split}
\end{equation}

\end{widetext}
where $\Delta E_{N_\alpha} = UN_\alpha + \alpha_{\rm g}e V_{\rm g}$.
In a last step we insert $g_{\alpha \tau}(\epsilon)$, see Eq.~\eqref{eq:dos} in the
main text, and the remaining
integral can be done by using the following
identities:
\begin{equation}
f^+(x) f^-(y) = b^+(x-y)\big(  f^+(y) -f^+(x)  \big),
\end{equation}
\begin{equation}
\int_{-\infty}^{\infty}  dx ~\big( f^+(x)- f^+(x+\omega)  \big)= \omega,
\end{equation}
\begin{widetext}
\begin{equation}\label{eq:prod_3_fermi}
\begin{split}
& \int_{-\infty}^{\infty}  dx ~ f^+(x+a) f^-(x+b) f^-(x+c)  =
\int_{-\infty}^{\infty}  dx ~  b^+(a-b)\bigg( f^+(x+b) - f^+(x+a) \bigg)
f^-(x+c)  =
\\  = &
 b^+(a-b) \bigg\{
b^+(b-c) ~ \int_{-\infty}^{\infty}  dx ~  \bigg( f^+(x+c) - f^+(x+b)  \bigg)
- b^+(a-c)~    \int_{-\infty}^{\infty}  dx ~  \bigg( f^+(x+c)- f^+(x+a)  \bigg)
\bigg\}
\\=&
 b^+(a-b)\bigg(  F(b-c) - F(a-c)
 \bigg)
\end{split}
\end{equation}
\end{widetext}
 $b^\pm(x)$ and $F(x)$ are defined in the main text just below  Eq.~\eqref{eq:rates_R}.
Using these identities yields the final result

\begin{equation}\label{eq:master_equation_final_app}
\begin{split}
\tr{\rm S} \big\{ & \mathcal{P}_{N_{\alpha}}   \dot{\hat\rho}^\alpha_{\infty}
\big\}
 =0 \\
 =  \sum_{ \eta \sigma } &
 \bigg\{
-
\Gamma^{N_{\alpha} \to N_{\alpha}-1  }_{\eta\alpha \sigma}~
P_{N_{\alpha}}
-
  \Gamma^{N_{\alpha} \to N_{\alpha} + 1  }_{\eta\alpha \sigma}~
  P_{N_{\alpha}}  \\
& +
\Gamma^{N_{\alpha}-1 \to N  }_{\eta\alpha \sigma}~
\, P_{N_{\alpha}-1}
+
 \Gamma^{N_{\alpha} +1 \to N }_{\eta\alpha \sigma}~
\,  P_{N_{\alpha} +1}
 \bigg\} .
\end{split}
\end{equation}

\section{Current}
Finally we briefly outline the derivation of the current formula. The current is
defined as
\begin{equation}
 I_\eta = e\frac{d}{dt}\braket{\hat N_\eta}(t).
\end{equation}
In the interaction picture the total particle number operator  of
lead $\eta $, $\hat N_\eta$, is not evolving in time  since it
commutes with the unperturbed part of the Hamiltonian.
Therefore, the current reads
\begin{widetext}
\begin{equation}\label{eq:current1}
\begin{split}
 I_\eta &= e \tr{\rm S+L} \bigg \{
\hat N_\eta   \frac{d}{dt}  \hat \rho_{\rm I}(t)
 \bigg \} \\
&=
-\ihbar \tr{\rm S+L} \bigg \{
\hat N_\eta    \big[ \hat H_{\rm T,I}(t), \hat \rho_{\rm I}(0) \big]
 \bigg \}
- \frac{1}{\hbar^2}  \int_0^{t} dt'~ \tr{\rm S+L} \bigg \{
\hat N_\eta     \bigg[  \hat H_{\rm T,I}(t),  \big[ \hat
H_{\rm T,I}(t'), \hat \rho_{\rm I}(t') \big]  \bigg]
 \bigg \}
\\
\end{split}
\end{equation}
\end{widetext}
where we expand $\frac{d}{dt}\hat \rho_{\rm I}(t)$ up to second order in $\hat H_{\rm T}$.
The first term of  Eq.~(\ref{eq:current1}) vanishes
since only a odd number of operators appear in the trace. In the second term
we replace $\hat \rho_{\rm I}(t')\to \rho_{\rm I}(t) $. Exploiting further the cyclic
invariance of the trace we find
 \begin{widetext}
\begin{equation}\label{eq:current2}
\begin{split}
 I_\eta &=
- \frac{e}{\hbar^2}  \int_0^{t} dt'~ \tr{} \bigg \{
   \bigg[ \big[ \hat N_\eta  , \hat H_{\rm T,I}(t)  \big] ,   \hat
H_{\rm T,I}(t')\bigg] \hat \rho_{\rm I}(t)
 \bigg \}
\\
&= - \frac{2e}{\hbar^2} \mathrm{Re}\bigg( \int_0^{t} dt'~ \tr{\rm S+L} \bigg \{
    \big[ \hat N_\eta  , \hat H_{\rm T,I}(t)  \big]   \hat
H_{\rm T,I}(t') \hat \rho_{\rm I}(t)
 \bigg \} \bigg)
\end{split}
\end{equation}
\end{widetext}
In the last step we exploited the anti-hermiticity of $\big[ \hat
N_\eta , \hat H_{\rm T,I}(t)  \big]$.
Following the same steps as in the derivation of the master equation, one can
identify  the rates, and one finds the well known expression of the current
\begin{equation}\label{eq:current_final_app}
\begin{split}
 I_\eta =&
- e \sum_{\alpha \sigma} \sum_{N_\alpha } \bigg\{  \Gamma^{N_\alpha\to N_\alpha
+1 }_{\eta\alpha \sigma}   P_{N_\alpha}  - \Gamma^{N_\alpha  \to N_\alpha -1
 }_{\eta\alpha \sigma}  P_{N_\alpha}      \bigg\}.
\end{split}
\end{equation}

\section{Calculation of the differential conductance}

Differentiating Eq.~(\ref{eq:IP_linear}) with respect to $V_{\rm b}$ and inserting
the definition of Eq.~(\ref{eq:def_slope_rates}) yields the differential
conductance of the first plateau:
\begin{equation}\label{eq:dIP_linear}
\frac{dI_{\rm D}^P }{d (V_{\rm b})}= \frac{2\pi e^2}{\hbar} D_0 \tilde{g}_\down \frac{(1-p)}{2}
\frac{\alpha_{\rm D} |t_{\rm D}|^2 \alpha_{\rm S} |t_{\rm S}|^2  }{ \alpha_D |t_{\rm D}|^2 - \alpha_{\rm S}
|t_{\rm S}|^2  }.
\end{equation}
To calculate the differential conductance  at this point we differentiate
\Eq{\ref{eq:IQ_linear}} with respect the bias voltage and find
\begin{equation}\label{eq:dIQ_linear}
\frac{dI_D^{(Q)} }{dV_{\rm b}} = -e^2\frac{ \alpha \gamma + 2\beta \gamma V_{\rm b} +
\beta V_{\rm b}^2 }{(\gamma + \delta V_{\rm b})^2},
\end{equation}
where we defined
$ \alpha = -\mc{B}_{\rm S\down}\mc{A}_{\rm D\up}  $,
$\beta = -\mc{B}_{\rm S\down} ( \mc{B}_{\rm D\up} + \mc{B}_{\rm D\down}  )  $,
$\gamma = \mc{A}_{\rm D\up} $, and
$\delta =  -\mc{B}_{\rm S\down} + \mc{B}_{\rm D\up} + \mc{B}_{\rm D\down}$.
In order to find the value of the differential conductance plateau we have to
consider the high bias limit and we find
\begin{equation}
\lim_{V_{\rm b} \to \infty}  \frac{dI_S^{(Q)} }{dV_{\rm b}} = -e^2\frac{\beta}{\delta} =
e^2\frac{ \mc{B}_{\rm S\down} ( \mc{B}_{\rm D\up} + \mc{B}_{\rm D\down}  )}{-\mc{B}_{\rm S\down} +
\mc{B}_{\rm D\up}  + \mc{B}_{\rm D\down}}.
\end{equation}
Inserting back the physical constants we find
\begin{equation}
\begin{split}
&\lim_{V_{\rm b} \to \infty}  \frac{dI_D^{(Q)} }{dV_{\rm b}} =
e^2\frac{2\pi}{\hbar} D_0 \tilde{g}_\down \frac{(1-p)}{2} \alpha_{\rm S} |t_{\rm S}|^2
\\
&\times \frac{\alpha_D |t_{\rm D}|^2 \big( (1+p) \tilde{g}_\up  + (1-p) \tilde{g}_\down   \big) }
{-(1-p)\tilde{g}_\down \alpha_{\rm S} |t_{\rm S}|^2 + \big( (1+p) \tilde{g}_\up  + (1-p) \tilde{g}_\down
\big)  \alpha_D |t_{\rm D}|^2  }.
\end{split}
\end{equation}

~~
~

% \bibliography{bibCB}
\bibliography{bibliography}

\begin{thebibliography}{18}
\expandafter\ifx\csname natexlab\endcsname\relax\def\natexlab#1{#1}\fi
\expandafter\ifx\csname bibnamefont\endcsname\relax
  \def\bibnamefont#1{#1}\fi
\expandafter\ifx\csname bibfnamefont\endcsname\relax
  \def\bibfnamefont#1{#1}\fi
\expandafter\ifx\csname citenamefont\endcsname\relax
  \def\citenamefont#1{#1}\fi
\expandafter\ifx\csname url\endcsname\relax
  \def\url#1{\texttt{#1}}\fi
\expandafter\ifx\csname urlprefix\endcsname\relax\def\urlprefix{URL }\fi
\providecommand{\bibinfo}[2]{#2}
\providecommand{\eprint}[2][]{\url{#2}}

\bibitem[{\citenamefont{Ohno et~al.}(1996)\citenamefont{Ohno, Shen, Matsukura,
  Oiwa, Endo, Katsumoto, and Iye}}]{Ohno96}
\bibinfo{author}{\bibfnamefont{H.}~\bibnamefont{Ohno}},
  \bibinfo{author}{\bibfnamefont{A.}~\bibnamefont{Shen}},
  \bibinfo{author}{\bibfnamefont{F.}~\bibnamefont{Matsukura}},
  \bibinfo{author}{\bibfnamefont{A.}~\bibnamefont{Oiwa}},
  \bibinfo{author}{\bibfnamefont{A.}~\bibnamefont{Endo}},
  \bibinfo{author}{\bibfnamefont{S.}~\bibnamefont{Katsumoto}},
  \bibnamefont{and} \bibinfo{author}{\bibfnamefont{Y.}~\bibnamefont{Iye}},
  \bibinfo{journal}{Applied Physics Letters} \textbf{\bibinfo{volume}{69}},
  \bibinfo{pages}{363} (\bibinfo{year}{1996}).

\bibitem[{\citenamefont{Dietl and Ohno}(2014)}]{DietlOhno14}
\bibinfo{author}{\bibfnamefont{T.}~\bibnamefont{Dietl}} \bibnamefont{and}
  \bibinfo{author}{\bibfnamefont{H.}~\bibnamefont{Ohno}},
  \bibinfo{journal}{Rev. Mod. Phys.} \textbf{\bibinfo{volume}{86}},
  \bibinfo{pages}{187} (\bibinfo{year}{2014}).

\bibitem[{\citenamefont{Jungwirth et~al.}(2006)\citenamefont{Jungwirth, Sinova,
  Ma\ifmmode~\check{s}\else \v{s}\fi{}ek, Ku\ifmmode~\check{c}\else
  \v{c}\fi{}era, and MacDonald}}]{Jungwirth2006}
\bibinfo{author}{\bibfnamefont{T.}~\bibnamefont{Jungwirth}},
  \bibinfo{author}{\bibfnamefont{J.}~\bibnamefont{Sinova}},
  \bibinfo{author}{\bibfnamefont{J.}~\bibnamefont{Ma\ifmmode~\check{s}\else
  \v{s}\fi{}ek}},
  \bibinfo{author}{\bibfnamefont{J.}~\bibnamefont{Ku\ifmmode~\check{c}\else
  \v{c}\fi{}era}}, \bibnamefont{and} \bibinfo{author}{\bibfnamefont{A.~H.}
  \bibnamefont{MacDonald}}, \bibinfo{journal}{Rev. Mod. Phys.}
  \textbf{\bibinfo{volume}{78}}, \bibinfo{pages}{809} (\bibinfo{year}{2006}).

\bibitem[{\citenamefont{Sato et~al.}(2010)\citenamefont{Sato, Bergqvist,
  Kudrnovsk\'y, Dederichs, Eriksson, Turek, Sanyal, Bouzerar, Katayama-Yoshida,
  Dinh et~al.}}]{Sato2010}
\bibinfo{author}{\bibfnamefont{K.}~\bibnamefont{Sato}},
  \bibinfo{author}{\bibfnamefont{L.}~\bibnamefont{Bergqvist}},
  \bibinfo{author}{\bibfnamefont{J.}~\bibnamefont{Kudrnovsk\'y}},
  \bibinfo{author}{\bibfnamefont{P.~H.} \bibnamefont{Dederichs}},
  \bibinfo{author}{\bibfnamefont{O.}~\bibnamefont{Eriksson}},
  \bibinfo{author}{\bibfnamefont{I.}~\bibnamefont{Turek}},
  \bibinfo{author}{\bibfnamefont{B.}~\bibnamefont{Sanyal}},
  \bibinfo{author}{\bibfnamefont{G.}~\bibnamefont{Bouzerar}},
  \bibinfo{author}{\bibfnamefont{H.}~\bibnamefont{Katayama-Yoshida}},
  \bibinfo{author}{\bibfnamefont{V.~A.} \bibnamefont{Dinh}},
  \bibnamefont{et~al.}, \bibinfo{journal}{Rev. Mod. Phys.}
  \textbf{\bibinfo{volume}{82}}, \bibinfo{pages}{1633} (\bibinfo{year}{2010}).

\bibitem[{\citenamefont{R\"uster et~al.}(2003)\citenamefont{R\"uster, Borzenko,
  Gould, Schmidt, Molenkamp, Liu, Wojtowicz, Furdyna, Yu, and
  Flatté}}]{Ruster2003}
\bibinfo{author}{\bibfnamefont{C.}~\bibnamefont{R\"uster}},
  \bibinfo{author}{\bibfnamefont{T.}~\bibnamefont{Borzenko}},
  \bibinfo{author}{\bibfnamefont{C.}~\bibnamefont{Gould}},
  \bibinfo{author}{\bibfnamefont{G.}~\bibnamefont{Schmidt}},
  \bibinfo{author}{\bibfnamefont{L.}~\bibnamefont{Molenkamp}},
  \bibinfo{author}{\bibfnamefont{X.}~\bibnamefont{Liu}},
  \bibinfo{author}{\bibfnamefont{T.}~\bibnamefont{Wojtowicz}},
  \bibinfo{author}{\bibfnamefont{J.}~\bibnamefont{Furdyna}},
  \bibinfo{author}{\bibfnamefont{Z.}~\bibnamefont{Yu}}, \bibnamefont{and}
  \bibinfo{author}{\bibfnamefont{M.}~\bibnamefont{Flatté}},
  \bibinfo{journal}{Phys. Rev. Lett.} \textbf{\bibinfo{volume}{91}},
  \bibinfo{pages}{216602} (\bibinfo{year}{2003}).

\bibitem[{\citenamefont{Giddings et~al.}(2005)\citenamefont{Giddings, Khalid,
  Jungwirth, Wunderlich, Yasin, Campion, Edmonds, Sinova, Ito, Wang
  et~al.}}]{Giddings2005}
\bibinfo{author}{\bibfnamefont{A.}~\bibnamefont{Giddings}},
  \bibinfo{author}{\bibfnamefont{M.}~\bibnamefont{Khalid}},
  \bibinfo{author}{\bibfnamefont{T.}~\bibnamefont{Jungwirth}},
  \bibinfo{author}{\bibfnamefont{J.}~\bibnamefont{Wunderlich}},
  \bibinfo{author}{\bibfnamefont{S.}~\bibnamefont{Yasin}},
  \bibinfo{author}{\bibfnamefont{R.}~\bibnamefont{Campion}},
  \bibinfo{author}{\bibfnamefont{K.}~\bibnamefont{Edmonds}},
  \bibinfo{author}{\bibfnamefont{J.}~\bibnamefont{Sinova}},
  \bibinfo{author}{\bibfnamefont{K.}~\bibnamefont{Ito}},
  \bibinfo{author}{\bibfnamefont{K.-Y.} \bibnamefont{Wang}},
  \bibnamefont{et~al.}, \bibinfo{journal}{Phys. Rev. Lett.}
  \textbf{\bibinfo{volume}{94}} (\bibinfo{year}{2005}).

\bibitem[{\citenamefont{Schlapps et~al.}(2006)\citenamefont{Schlapps, Doeppe,
  Wagner, Reinwald, Wegscheider, and Weiss}}]{Schlapps06}
\bibinfo{author}{\bibfnamefont{M.}~\bibnamefont{Schlapps}},
  \bibinfo{author}{\bibfnamefont{M.}~\bibnamefont{Doeppe}},
  \bibinfo{author}{\bibfnamefont{K.}~\bibnamefont{Wagner}},
  \bibinfo{author}{\bibfnamefont{M.}~\bibnamefont{Reinwald}},
  \bibinfo{author}{\bibfnamefont{W.}~\bibnamefont{Wegscheider}},
  \bibnamefont{and} \bibinfo{author}{\bibfnamefont{D.}~\bibnamefont{Weiss}},
  \bibinfo{journal}{physica status solidi (a)} \textbf{\bibinfo{volume}{203}},
  \bibinfo{pages}{3597} (\bibinfo{year}{2006}), ISSN \bibinfo{issn}{1862-6319}.

\bibitem[{\citenamefont{Ciorga et~al.}(2007)\citenamefont{Ciorga, Schlapps,
  Einwanger, Geissler, Sadowski, Wegscheider, and Weiss}}]{Ciorga2007}
\bibinfo{author}{\bibfnamefont{M.}~\bibnamefont{Ciorga}},
  \bibinfo{author}{\bibfnamefont{M.}~\bibnamefont{Schlapps}},
  \bibinfo{author}{\bibfnamefont{A.}~\bibnamefont{Einwanger}},
  \bibinfo{author}{\bibfnamefont{S.}~\bibnamefont{Geissler}},
  \bibinfo{author}{\bibfnamefont{J.}~\bibnamefont{Sadowski}},
  \bibinfo{author}{\bibfnamefont{W.}~\bibnamefont{Wegscheider}},
  \bibnamefont{and} \bibinfo{author}{\bibfnamefont{D.}~\bibnamefont{Weiss}},
  \bibinfo{journal}{New J. Phys.} \textbf{\bibinfo{volume}{9}},
  \bibinfo{pages}{351} (\bibinfo{year}{2007}).

\bibitem[{\citenamefont{Pappert et~al.}(2007)\citenamefont{Pappert, Huempfner,
  Gould, Wenisch, Brunner, Schmidt, and Molenkamp}}]{Pappert2007}
\bibinfo{author}{\bibfnamefont{K.}~\bibnamefont{Pappert}},
  \bibinfo{author}{\bibfnamefont{S.}~\bibnamefont{Huempfner}},
  \bibinfo{author}{\bibfnamefont{C.}~\bibnamefont{Gould}},
  \bibinfo{author}{\bibfnamefont{J.}~\bibnamefont{Wenisch}},
  \bibinfo{author}{\bibfnamefont{K.}~\bibnamefont{Brunner}},
  \bibinfo{author}{\bibfnamefont{G.}~\bibnamefont{Schmidt}}, \bibnamefont{and}
  \bibinfo{author}{\bibfnamefont{L.~W.} \bibnamefont{Molenkamp}},
  \bibinfo{journal}{Nature Physics} \textbf{\bibinfo{volume}{3}},
  \bibinfo{pages}{573} (\bibinfo{year}{2007}).

\bibitem[{\citenamefont{Wunderlich et~al.}(2006)\citenamefont{Wunderlich,
  Jungwirth, Kaestner, Irvine, Shick, Stone, Wang, Rana, Giddings, Foxon
  et~al.}}]{Wunderlich2006}
\bibinfo{author}{\bibfnamefont{J.}~\bibnamefont{Wunderlich}},
  \bibinfo{author}{\bibfnamefont{T.}~\bibnamefont{Jungwirth}},
  \bibinfo{author}{\bibfnamefont{B.}~\bibnamefont{Kaestner}},
  \bibinfo{author}{\bibfnamefont{A.}~\bibnamefont{Irvine}},
  \bibinfo{author}{\bibfnamefont{A.}~\bibnamefont{Shick}},
  \bibinfo{author}{\bibfnamefont{N.}~\bibnamefont{Stone}},
  \bibinfo{author}{\bibfnamefont{K.-Y.} \bibnamefont{Wang}},
  \bibinfo{author}{\bibfnamefont{U.}~\bibnamefont{Rana}},
  \bibinfo{author}{\bibfnamefont{A.}~\bibnamefont{Giddings}},
  \bibinfo{author}{\bibfnamefont{C.}~\bibnamefont{Foxon}},
  \bibnamefont{et~al.}, \bibinfo{journal}{Phys. Rev. Lett.}
  \textbf{\bibinfo{volume}{97}} (\bibinfo{year}{2006}).

\bibitem[{\citenamefont{Schlapps et~al.}(2009)\citenamefont{Schlapps, Lermer,
  Geissler, Neumaier, Sadowski, Schuh, Wegscheider, and Weiss}}]{Schlapps2009}
\bibinfo{author}{\bibfnamefont{M.}~\bibnamefont{Schlapps}},
  \bibinfo{author}{\bibfnamefont{T.}~\bibnamefont{Lermer}},
  \bibinfo{author}{\bibfnamefont{S.}~\bibnamefont{Geissler}},
  \bibinfo{author}{\bibfnamefont{D.}~\bibnamefont{Neumaier}},
  \bibinfo{author}{\bibfnamefont{J.}~\bibnamefont{Sadowski}},
  \bibinfo{author}{\bibfnamefont{D.}~\bibnamefont{Schuh}},
  \bibinfo{author}{\bibfnamefont{W.}~\bibnamefont{Wegscheider}},
  \bibnamefont{and} \bibinfo{author}{\bibfnamefont{D.}~\bibnamefont{Weiss}},
  \bibinfo{journal}{Phys. Rev. B} \textbf{\bibinfo{volume}{80}},
  \bibinfo{pages}{125330} (\bibinfo{year}{2009}).

\bibitem[{\citenamefont{Averin and Likharev}(1986)}]{Averin1986}
\bibinfo{author}{\bibfnamefont{D.~V.} \bibnamefont{Averin}} \bibnamefont{and}
  \bibinfo{author}{\bibfnamefont{K.}~\bibnamefont{Likharev}},
  \bibinfo{journal}{J. Low Temp. Phys.} p. \bibinfo{pages}{345}
  (\bibinfo{year}{1986}).

\bibitem[{\citenamefont{Averin and Likharev}(1991)}]{Averin1991}
\bibinfo{author}{\bibfnamefont{D.~V.} \bibnamefont{Averin}} \bibnamefont{and}
  \bibinfo{author}{\bibfnamefont{K.~K.} \bibnamefont{Likharev}},
  \emph{\bibinfo{title}{Mesoscopic Phenomena in Solids}}
  (\bibinfo{publisher}{Elsevier Science, Amsterdam}, \bibinfo{year}{1991}).

\bibitem[{\citenamefont{Grabert}(1991)}]{Grabert1991}
\bibinfo{author}{\bibfnamefont{H.}~\bibnamefont{Grabert}},
  \bibinfo{journal}{Zeitschrift F\"ur Phys. B Condens. Matter}
  \textbf{\bibinfo{volume}{85}}, \bibinfo{pages}{319} (\bibinfo{year}{1991}).

\bibitem[{\citenamefont{Grabert and Devoret}(1992)}]{Grabert1992}
\bibinfo{editor}{\bibfnamefont{H.}~\bibnamefont{Grabert}} \bibnamefont{and}
  \bibinfo{editor}{\bibfnamefont{M.~H.} \bibnamefont{Devoret}}, eds.,
  \emph{\bibinfo{title}{Sigle Charge Tunneling}}, NATO ASI Series
  (\bibinfo{publisher}{Springer US}, \bibinfo{year}{1992}).

\bibitem[{\citenamefont{Sohn et~al.}(1997)\citenamefont{Sohn, Kouwenhoven, and
  Sch\"{o}n}}]{Kouwenhoven1997}
\bibinfo{editor}{\bibfnamefont{L.}~\bibnamefont{Sohn}},
  \bibinfo{editor}{\bibfnamefont{L.}~\bibnamefont{Kouwenhoven}},
  \bibnamefont{and}
  \bibinfo{editor}{\bibfnamefont{G.}~\bibnamefont{Sch\"{o}n}}, eds.,
  \emph{\bibinfo{title}{Mesoscopic Electron Transport}}, NATO ASI Series
  (\bibinfo{publisher}{Kluwer}, \bibinfo{year}{1997}).

\bibitem[{\citenamefont{Barna\'{s} and Weymann}(2008)}]{Barnas2008}
\bibinfo{author}{\bibfnamefont{J.}~\bibnamefont{Barna\'{s}}} \bibnamefont{and}
  \bibinfo{author}{\bibfnamefont{I.}~\bibnamefont{Weymann}},
  \bibinfo{journal}{J. Phys.: Condens. Matter} \textbf{\bibinfo{volume}{20}},
  \bibinfo{pages}{423202} (\bibinfo{year}{2008}).

\bibitem[{\citenamefont{Edmonds et~al.}(2004)\citenamefont{Edmonds,
  Bogus\-lawski, Wang, Campion, Novikov, Farley, Gallagher, Foxon, Sawicki,
  Dietl et~al.}}]{Edmonds2004}
\bibinfo{author}{\bibfnamefont{K.}~\bibnamefont{Edmonds}},
  \bibinfo{author}{\bibfnamefont{P.}~\bibnamefont{Bogus\-lawski}},
  \bibinfo{author}{\bibfnamefont{K.}~\bibnamefont{Wang}},
  \bibinfo{author}{\bibfnamefont{R.}~\bibnamefont{Campion}},
  \bibinfo{author}{\bibfnamefont{S.}~\bibnamefont{Novikov}},
  \bibinfo{author}{\bibfnamefont{N.}~\bibnamefont{Farley}},
  \bibinfo{author}{\bibfnamefont{B.}~\bibnamefont{Gallagher}},
  \bibinfo{author}{\bibfnamefont{C.}~\bibnamefont{Foxon}},
  \bibinfo{author}{\bibfnamefont{M.}~\bibnamefont{Sawicki}},
  \bibinfo{author}{\bibfnamefont{T.}~\bibnamefont{Dietl}},
  \bibnamefont{et~al.}, \bibinfo{journal}{Phys. Rev. Lett.}
  \textbf{\bibinfo{volume}{92}}, \bibinfo{pages}{037201}
  (\bibinfo{year}{2004}).

\end{thebibliography}

\end{document}